\documentclass[twocolumn,fleqn]{autart} 

\usepackage{amssymb}
\usepackage{mathtools}
\usepackage{graphicx}
\usepackage[export]{adjustbox}
\usepackage[dvipsnames]{xcolor}
\usepackage{hyperref}
\usepackage{cancel}
\usepackage{verbatim} % includes comment
\usepackage{tikz}

\newcommand{\reals}{\mathbb{R}}
\newcommand{\transpose}{^\textrm{T}}
\newcommand{\continuous}{\mathcal{C}}
\newcommand{\argmin}{\operatorname*{arg\,min}}
\newcommand{\argmax}{\operatorname*{arg\,max}}
\newcommand{\interior}{\operatorname{int}}

\newcommand{\highdegree}{\mathcal{G}}
\renewcommand{\(}{\left(}
\renewcommand{\)}{\right)}
\renewcommand{\[}{\left[}
\renewcommand{\]}{\right]}

 \newcommand{\regularversion}[1]{\iffalse #1 \fi}
 \newcommand{\extendedversion}[1]{{#1}}
\begin{document}

\begin{frontmatter}

\title{Robust Control Barrier Functions under High Relative Degree and Input Constraints for Satellite Trajectories\thanksref{footnoteinfo}}

\thanks[footnoteinfo]{Part of this paper was {\color{black}presented at} the 2021 IEEE Conference on Decision and Control. \\\hspace*{7pt} Corresponding author J.~Breeden}

\author[UM]{Joseph Breeden}\ead{jbreeden@umich.edu},
\author[UMR]{Dimitra Panagou}\ead{dpanagou@umich.edu}

\address[UM]{Department of Aerospace Engineering, University of Michigan, MI, USA}
\address[UMR]{\color{black}Department of Robotics and Department of Aerospace Engineering, University of Michigan, MI, USA}

% Five to ten keywords,  chosen from the IFCA keyword list or with the help of the Automatica keyword wizard

\begin{abstract} % Abstract of not more than 200 words.
This paper presents methodologies for constructing Control Barrier Functions (CBFs) for nonlinear, control-affine systems, in the presence of input constraints and bounded disturbances.
More specifically, given a constraint function with high-relative-degree with respect to the system dynamics, the paper considers three methodologies, two for relative-degree 2 and one for higher relative-degrees, for 
{\color{black}creating CBFs whose zero sublevel sets are subsets of the constraint function's zero sublevel set.}
Three special forms of Robust CBFs (RCBFs) are developed as functions of the input constraints, system dynamics, and disturbance bounds, such that 
the resultant RCBF condition on the control input {\color{black}is always feasible for states in the RCBF zero sublevel set. The RCBF condition is then} enforced in a switched fashion, which {\color{black}allows the {\color{black}system} to operate {\color{black}safely without enforcing the RCBF condition} 
when far from the safe set boundary and allows tuning of how closely trajectories approach the safe set boundary.} 
The proposed methods are verified in simulations demonstrating the developed RCBFs in an asteroid flyby scenario for a satellite with low-thrust actuators\extendedversion{, and in asteroid proximity operations for a satellite with high-thrust actuators}.
\end{abstract}

\end{frontmatter}

%%%%%%%%%%%%%%%%%%%%%%%%%%%%%%%%%%%%%%%%%%%%%%%%%%%%%%%%%%%%%%%%%%%%%%%%%%%%%%%%
\section{Introduction}

This paper advances the recent theory of Control Barrier Functions (CBFs) to systems with higher relative-degree under input constraints and disturbances, and applies the results to control design for satellite trajectories. 
Currently, satellite trajectory design is generally the product of extensive optimizations for fuel and/or time consumption, completed long before a satellite is deployed.
As spacecraft venture further away from the Earth and attempt more complex mission objectives, there is a need for greater autonomy, and a subsequent need to ensure {\color{black}that autonomous trajectories meet various requirements, herein termed \textit{safety}. While our focus is on spacecraft, the following results are broadly applicable to systems with constraints of high-relative-degree.}

System safety is often formulated as an invariance problem for a set of safe states, referred to as the \textit{safe set}. 
In this paper, the safe set is defined as the zero sublevel set of some \textit{constraint function}, which is assumed to be of high-relative-degree with respect to the system dynamics, as is the case for satellite problems. We then 
design a CBF whose zero sublevel set, which we call the \textit{inner safe set}, is a subset of the safe set. 
Existing CBF theory then provides a sufficient condition, which we call the \textit{CBF condition}, on the control input that establishes forward invariance of the inner safe set (for an overview of CBFs, see \cite{CBF_Tutorial}). 
However, one limitation of this approach is that finding a valid CBF may be challenging in general. Thus, the main questions of this paper are: given a constraint function with relative-degree 2 (or higher in Section~\ref{sec:predictive_cbf}) {\color{black}and specified input constraints}, 1) how to determine a CBF whose inner safe set is a subset of the safe set and which is valid with respect to the input constraints, and 2) how to ensure invariance of the inner safe set in the presence of bounded disturbances.

Given this setup, several papers develop methods on constructing CBFs whose inner safe sets are equivalent to the safe set. Prior methods include constructing CBFs via compositions of the constraint function and its derivatives \cite{Ames_Journal}, backstepping \cite{Backstepping_CBF}, or {\color{black}feedback linearization \cite{Exponential_CBF,Control_Sharing,CDC20,choi2020reinforcement}}. 
However, the aforementioned papers all require that the set of allowable control inputs is $\reals^m$. Since the constraint function is of high-relative-degree, there may exist states in the safe set from which arbitrarily large control inputs are needed to render the system trajectories within the safe set. Therefore, if the set of allowable control inputs is bounded, there may be no admissible control input that keeps the system trajectories within the safe set, and hence the system could become unsafe (e.g., the satellite might fail to decelerate before colliding with an obstacle).
The work in \cite{Recursive_CBF} fixes this issue by specifying the inner safe set as a viability domain for the given system dynamics---i.e. a set inside the safe set that may be rendered forward invariant under input constraints---and provides parameters (class-$\mathcal{K}$ functions) that can be tuned to meet various input constraints. 
{\color{black}However},
finding such parameters is a non-trivial challenge, akin to finding a Lyapunov function.
{\color{black}Recently, such parameters have been found via learning from expert demonstration \cite{robey2021learning} and reinforcement learning \cite{jin2020neural}.}
The results of this paper will similarly specify a subset of the safe set that is controlled forward invariant but, {\color{black}unlike \cite{Recursive_CBF}, here we use first-order CBFs \cite{CBF_Tutorial} and we provide three principled methodologies to select} the analogue of {\color{black}the parameters in \cite{Recursive_CBF}} {\color{black}for certain classes of systems (namely, systems satisfying the theorem assumptions in Section~\ref{sec:new_cbfs})}.
{\color{black} These methodologies are based {\color{black}on} feedback linearization (distinct from \cite{Exponential_CBF}), potential energy functions, and model-predictive safety, respectively.}
The significance of these results and those in {\color{black}\cite{Recursive_CBF,robey2021learning,jin2020neural}} is that if a {\color{black}valid} CBF can be found {\color{black}via one of these methods}, then existence of a safe trajectory is guaranteed from any point in the inner safe set.

Other studies have taken advantage of special features of certain systems to satisfy input constraints when the constraint function is of high-relative-degree. The work in \cite{MultiRobotsJournal} develops a CBF specific to the double integrator system. Similarly, \cite{Integrator_CBF} develops a technique applicable to the $n$-integrator system, which is generalized to similar systems in \cite{CDC21}. {\color{black}The method in \cite[Sec. III-B]{CDC21} is then extended in Section~\ref{sec:abs_cbf} to be robust to disturbances while maintaining provable safety under input constraints, and in Section~\ref{sec:approach_func} to work with more general dynamics for which \cite[Eq. 17]{CDC21} is possibly zero}. Alternatively, one could use an additional CBF to limit the agent velocity to a certain domain, as in \cite{Ames_Journal,MultiRobotsJournal}. The example in \cite{First_CBF} instead takes advantage of the damping of the considered system to meet input constraints. {\color{black}Finally, the authors' prior work in \cite[Sec. III-A]{CDC21} and other authors \cite{Predictive_CBF,backup_controller} have considered CBFs that examine a system's trajectory forward in time under various assumptions to determine safety. The authors of \cite{backup_controller} consider the set of states reachable in fixed time from a pre-designated \textit{backup set}, but do not consider whether the predicted trajectories from the current state to the backup set are everywhere safe. The work in \cite{Predictive_CBF} improves upon this by examining the minimizer of a \textit{performance function} applied along the predicted trajectories, and \cite{CDC21} extends the method to work when the minimizer is not unique. Compared to these works, Section~\ref{sec:predictive_cbf} extends this strategy to be robust to disturbances while maintaining provable safety under input constraints, and provides tools for computing the CBF and its derivatives when analytic solutions are unavailable (see also \cite[Sec. V-A]{backup_controller}).}

Several papers have considered CBF robustness to disturbances in various senses.
Neglecting input constraints, an early result on CBF robustness in \cite{Robust_CBFs} shows that a bounded disturbance causes a bounded excursion outside the CBF zero sublevel set, and later authors showed {\color{black}that} this excursion can be tuned \cite{tunable_safety}. 
{\color{black}Recently, \cite{robust_hocbfs} extended this result to higher-order CBFs as in \cite{Recursive_CBF}.}
Safety under a bounded worst-case disturbance{\color{black}, as is considered in this paper,} is studied in \cite{garg2020robust,RobustCBFsOld,Cortez}, while probability of safety using a similar approach with a stochastic disturbance is studied in \cite{Prajna2007}. In multi-agent systems, dynamic couplings between agents that act independently of each other can also be considered disturbances, and robustness to such effects are treated similarly in \cite{Sampled_Data_Adversarial,Conflicting_STL}. In all of these papers, it is assumed that the system has sufficient control authority to counteract these disturbances. However, satisfying this assumption is nontrivial, and is a requirement of the methods in Section~\ref{sec:new_cbfs}.

{\color{black}Finally, one objective of this paper is to {\color{black}place fewer restrictions on closed-loop trajectories by applying safety criteria only near the boundary of the inner safe set.} 
This was accomplished in \cite{robust_hocbfs} by designating strict subsets of the safe set termed ``performance-critical regions'' where safety was guaranteed without enforcing the CBF condition. However, this relaxation required expanding the control set to $\reals^m$. For systems subject to many CBFs simultaneously,} the work in \cite{CDC20,MultiRobotsJournal} simplifies control input calculation {\color{black}in a similar manner} by breaking the state space into regions where only a few CBFs are actively applied, though this introduces potential issues with non-uniqueness of system solutions. These issues are fixed in \cite{Glotfelter2} by relaxing the system to a differential inclusion{\color{black}.} 
A similar approach using products of CBFs is described in \cite{Additive_CBFs}. Expanding upon these approaches, this paper introduces a hysteresis-switching approach inspired by \cite{Glotfelter2} and \cite{First_CBF} that {\color{black}relaxes the CBF condition in the interior of the inner safe set while still provably guaranteeing safety in the presence of input constraints. Such hysteresis-switching removes the need for} differential inclusions, and thus prevents chattering control inputs that may not be feasible on real actuators. {\color{black}This switching approach also motivates a special choice for the class-$\mathcal{K}$ function that is left as a free tuning parameter in all of the theorems in Section~\ref{sec:new_cbfs}. We show in simulation how the proposed choice allows one to directly tune how closely trajectories approach the boundary of the safe set.}

In summary, the contributions of this paper are:
\begin{enumerate}
    \item three strategies for generating CBFs from high-relative-degree constraint functions in the presence of input constraints and bounded matched and unmatched disturbances simultaneously (Section~\ref{sec:new_cbfs});
    \item a switching method for {\color{black}relaxing the CBF condition in the interior of the inner safe set and tuning how closely trajectories approach the boundary of the inner safe set} (Section~\ref{sec:switching}); and
    \item specializations of the above CBFs to deep-space trajectory applications (Section~\ref{sec:spacecraft_functions}).
\end{enumerate}

\section{Preliminaries} \label{sec:prelims}

\subsection{Notation}

Given a time domain $\mathcal{D}\subseteq\reals$ and spatial domain $\mathcal{X}\subseteq\reals^n$ and a function $\eta:\mathcal{D}\times\mathcal{X}\rightarrow\reals$, denoted $\eta(t,x)$, let $\partial_t \eta$ denote the partial derivative with respect to the first variable, $t$. Let $\nabla \eta$ denote the gradient with respect to the second variable, $x$. Let $\frac{d}{dt}\eta$ denote the total derivative of $\eta$ in time, $\frac{d}{dt}\eta = \partial_t \eta + \nabla \eta \frac{dx}{dt}$. For brevity, let $\dot{\eta}$, $\ddot{\eta}$, and $\eta^{(r)}$ denote the first, second, and $r$th total derivative in time. Denote the derivative of a function $\kappa:\reals\rightarrow\reals$ as $\kappa'$ and its inverse (if it exists) as $\kappa^{-1}$. Additional derivative notation is introduced in Section~\ref{sec:predictive_cbf} to prevent confusion in that section. Let $I$ denote the identity matrix. Let $\| \cdot \|$ denote the 2-norm and $\| \cdot \|_\infty$ denote the $\infty$-norm. Let $\mathcal{C}^r$ be the set of functions that are $r$-times continuously differentiable in all arguments. A function {\color{black}$\alpha:\reals_{\geq 0}\rightarrow\reals_{\geq 0}$} is said to belong to {\color{black}class-$\mathcal{K}$}, denoted $\alpha\in\mathcal{K}$, if it is strictly increasing and $\alpha(0) = 0$. Given a set $\mathcal{S}$, let $\partial\mathcal{S}$ denote the boundary of $\mathcal{S}$. 

\subsection{Model and Problem}

Consider the time-varying control-affine model
\begin{equation}
    \dot{x} = \underbrace{f(t,x) + g(t,x) (u+w_u) + w_x}_{= F(t,x,u,w_u,w_x)} \,, \label{eq:model}
\end{equation}
with time $t\in \mathcal{D}=[t_0,{\color{black}\infty)}$, state $x\in\reals^n$, control input $u\in \mathcal U \subset \reals^m$ where $\mathcal{U}$ is compact, unknown disturbances $w_u \in \reals^m$ and $w_x\in\reals^n$ that are continuous in time, and functions $f:\mathcal D\times\reals^n\rightarrow\reals^n$ and $g:\mathcal D\times\reals^n\rightarrow\reals^{n\times m}$ that are piecewise continuous in $t$ and locally Lipschitz continuous in $x$. Let $w_u$ and $w_x$ be bounded as $\|w_u\| \leq w_{u,\textrm{max}}$ and $\|w_x\| \leq w_{x,\textrm{max}}$ for some $w_{u,\textrm{max}},w_{x,\textrm{max}} \in \reals_{\geq 0}$, and define the set of allowable disturbances $\mathcal{W} \triangleq \{ w_u \in \reals^m \mid \|w_u\|\leq w_{u,\textrm{max}} \} \times \{ w_x \in \reals^n \mid \|{\color{black}w_x}\|\leq w_{x,\textrm{max}}\}$. 
{\color{black}Assume a unique solution to \eqref{eq:model} exists for all $t\in\mathcal{D}$.}
Given dynamics \eqref{eq:model}, a function $\eta : \mathcal D\times\reals^n \rightarrow \reals$ is said to be of relative-degree $r$ if it is $r$-times total differentiable in time and $\eta^{(r)}$ is the lowest order derivative in which $u$ {\color{black}and $w_u$} appear explicitly. Denote the set of all relative-degree $r$ functions as $\mathcal{G}^r$.

Let $h : \mathcal D\times\reals^n \rightarrow \reals$, $h \in \highdegree^r$, denote the \textit{constraint function}, and define a safe set $\mathcal{S}$ as
\begin{equation}
    \mathcal S(t) \triangleq \{x\in\reals^n\mid h(t,x) \leq 0\} \,, \label{eq:safe_set}
\end{equation}
where we will henceforth drop the argument $t$ for compactness. Also, for compactness, denote the safe set across time as $\mathcal{T} \triangleq \{ (t,x) \in \reals\times\reals^n \mid t\in \mathcal{D}, x \in \mathcal S(t) \}$.

\begin{figure}
    \centering
    \includegraphics[width=0.98\columnwidth]{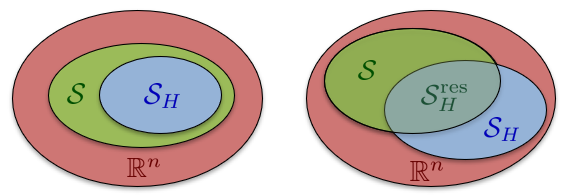}
    \caption{\color{black}Given a safe set $\mathcal{S} \subset \reals^n$ as in \eqref{eq:safe_set} and input constraints $\mathcal{U}$, this paper presents methods of finding viability domains $\mathcal{S}_H$ as in \eqref{eq:inner_safe_set} that are subsets of the safe set (left). In certain cases (Theorems~\ref{thm:constant_noise_cbf}-\ref{thm:variable_cbf}), the presented forms of $H$ cause the set $\mathcal{S}_H$ to include unsafe states (see also Fig.~\ref{fig:set_illustration}), so we introduce the set $\mathcal{S}_H^\textrm{res}$ in \eqref{eq:restricted_safe_set} (right). If $H$ is a RCBF as in Definition~\ref{def:rcbf}, then $\mathcal{S}_H$ (or $\mathcal{S}_H^\textrm{res}$ in {\color{black}Theorems~\ref{thm:constant_noise_cbf}-\ref{thm:variable_cbf}}) can be rendered forward invariant under any disturbances $(w_u,w_x)\in\mathcal{W}$ while always satisfying the input constraints $\mathcal{U}$.}
    \label{fig:sets_fig}
\end{figure}

This work is devoted to developing methods for rendering the state trajectory always inside the safe set $\mathcal S$ {\color{black}in the presence of any allowable disturbances $(w_u,w_x)\in\mathcal{W}$.} %, specifically by rendering a subset of $\mathcal S$ forward invariant. 
We will do this by constructing functions $H:\mathcal{D}\times\reals^n\rightarrow\reals$ 
{\color{black}that}
generate sets of the form
\begin{gather}
    \mathcal S_H(t) \triangleq \{ x \in \reals^n \mid H(t,x) \leq 0 \} \,,  \label{eq:inner_safe_set} \\
    \color{black}\mathcal S_H^{\textrm{\textnormal{res}}}(t) \triangleq \{ x \in \reals^n \mid H(t,x) \leq 0 \textrm{\textnormal{ and }} h(t,x) \leq 0 \} {\color{black}\,,} \label{eq:restricted_safe_set}
\end{gather}
{\color{black}visualized in Fig.~\ref{fig:sets_fig}.} We refer to the set $\mathcal S_H$ as an \textit{inner safe set}{\color{black}, and to the set $\mathcal{S}_H^\textrm{res}$ as a \textit{restricted safe set}. Note that if $H(t,x) \geq h(t,x)$ for all $(t,x)\in\mathcal{D}\times\reals^n$, then $\mathcal{S}_H\equiv \mathcal{S}_H^\textrm{res}$.} A controller is said to render ${\color{black}\mathcal S_H^\textrm{res}}$ forward invariant, if given any $x(t_0) \in {\color{black}\mathcal S_H^\textrm{res}(t_0)}$, the closed-loop trajectory satisfies $x(t)\in{\color{black}\mathcal{S}_H^\textrm{res}(t)}, \forall t\in\mathcal{D}$. 
In general, {\color{black}there may exist points $x(t_0)\in \mathcal{S}(t_0)$, from which} we will not be able to render $\mathcal S$ forward invariant under \eqref{eq:model}. Nevertheless, {\color{black}if we can render the subset}
${\color{black}\mathcal S_H^\textrm{res} \subseteq \mathcal{S}}$ forward invariant, {\color{black}then} we {\color{black}can} ensure that the closed loop trajectories of \eqref{eq:model} are safe {\color{black}(i.e. always stay in $\mathcal{S}$)} for initial conditions lying in the set ${\color{black}\mathcal S_H^\textrm{res}}$. Thus, a crucial requirement is that $x(t_0)\in{\color{black}\mathcal{S}_H^\textrm{res}}(t_0)$, where $H$ is chosen from the strategies in Section~\ref{sec:new_cbfs}. We also define the {\color{black}domains $\mathcal T_H \triangleq\{(t,x)\in\reals\times\reals^n\mid t\in\mathcal{D},x\in\mathcal{S}_H(t)\}$ and $\mathcal T_H^\textrm{res} \triangleq\{(t,x)\in\reals\times\reals^n\mid t\in\mathcal{D},x\in\mathcal{S}_H^\textrm{res}(t)\}$ similar} to $\mathcal T$.

\subsection{Mathematical Background}

All the CBFs considered in this paper will be Zeroing Control Barrier Functions as in \cite{Ames_Journal} with a time-varying extension as in \cite{CBFs_STL}. The following definition of CBF, which does not consider the disturbances, is inspired by \cite[Def. 3]{CBFs_STL}, and has been adapted to the notation of this paper.

\begin{defn}%[{\cite[Def. 3]{CBFs_STL}}] 
\label{def:cbf_definition}
    For the system \eqref{eq:model} with $w_u\equiv 0$, $w_x\equiv 0$, a continuously differentiable function $H : \mathcal{D}\times\reals^n \rightarrow\reals$ is a \textnormal{control barrier function (CBF)} on a time-varying set $\mathcal{X}$ if 
    there exists a locally Lipschitz continuous function $\alpha \in \mathcal{K}$ such that for all $x\in\mathcal{X}(t), t\in \mathcal{D}$
    \begin{multline} \label{eq:cbf_definition}
        \inf_{u\in \mathcal U} {\color{black}\big[} \partial_t H(t,x) + \nabla H(t,x) {\color{black}\big(}f(t,x)+g(t,x)u {\color{black}\big) \big]} \\ \leq \alpha(-H(t,x)).
    \end{multline}
\end{defn}

Given a CBF on the set $\mathcal{S}_H$, a condition for {\color{black}forward} invariance of $\mathcal{S}_H$ %(and hence, safety of the system) 
is then established in \cite[Thm. 1]{CBFs_STL}, provided there are no disturbances. 
Part of the proof of \cite[Thm. 1]{CBFs_STL} requires the following lemma, which we will also use for extensions of this theorem in Section{\color{black}s~\ref{sec:new_cbfs}-}\ref{sec:switching}.

\begin{lem}[{\color{black}\cite[Lem. 2]{Glotfelter1}, \cite[Lem. 1]{CBFs_STL}}] \label{prior:nonpositive_lemma}
    Let $\alpha \in \mathcal{K}$ be locally Lipschitz continuous and $\eta :\mathcal{D}\rightarrow\reals$ be absolutely continuous. If $\eta(t_0) \leq 0$ and $\dot{\eta}(t) \leq \alpha(-\eta(t))$ {\color{black}for almost every} $t \in \mathcal{D}$, then $\eta(t) \leq 0, \forall t \in \mathcal{D}$.
\end{lem}

Next, to account for disturbances, we introduce the following definition, inspired by \cite{garg2020robust,RobustCBFsOld}.

\begin{defn} \label{def:rcbf}
    For the system \eqref{eq:model}, a continuously differentiable function $H:\mathcal{D}\times\reals^n\rightarrow\reals$ is a \textnormal{robust control barrier function (RCBF)} on a time-varying set $\mathcal{X}$ if there exists a locally Lipschitz continuous $\alpha\in\mathcal{K}$ such that $\forall x \in \mathcal X(t), t\in \mathcal{D}$,
    \begin{multline}
        \max_{(w_u,w_x)\in\mathcal{W}} \Big[ \inf_{u\in \mathcal U} {\color{black}\Big(}\partial_t H(t,x) +  \nabla H(t,x) {\color{black}\big[}f(t,x) \\ + g(t,x) (u+w_u) + w_x{\color{black}\big] \Big)} \Big] \leq \alpha(-H(t,x)) \,. \label{eq:rcbf_definition}
    \end{multline}
\end{defn}
Based on Definition~\ref{def:rcbf}, we also define for compactness
\begin{align}
    W(t, &x) \triangleq \max_{(w_u,w_x)\in\mathcal{W}} \nabla H(t,x) (g(t,x) w_u + w_x) \label{eq:max_error} \\ 
    &= ||\nabla H(t,x) g(t,x)|| w_{u,\textrm{max}} + ||\nabla H(t,x)|| w_{x,\textrm{max}} \,, \nonumber
\end{align}
where we will use both equivalent forms of $W(t,x)$ in \eqref{eq:max_error} depending on the setting.
The set of control inputs such that \eqref{eq:rcbf_definition} is satisfied is then
\begin{multline} \label{eq:valid_rcbf_control}
    \boldsymbol\mu_\textrm{rcbf} (t,x) \triangleq \{ u\in\mathcal{U} \mid \partial_t H(t,x) + \nabla H(t,x)(f(t,x) \\ +g(t,x)u) \leq \alpha(-H(t,x)) - W(t,x) \} \,.
\end{multline}
Note that since Definition~\ref{def:rcbf} considers the allowable control set $\mathcal{U}$, if $H$ is a RCBF on $\mathcal{X}$, then $\boldsymbol\mu_\textrm{rcbf}(t,x)$ is nonempty for all $x\in\mathcal{X}(t),t\in\mathcal{D}$. 
Given a RCBF on $\mathcal{S}_H$, we can establish {\color{black}forward} invariance of $\mathcal{S}_H$ using the following lemma.
\begin{lem} \label{prior:rcbf_invariance}
    For the system \eqref{eq:model}, if $H:\mathcal{D}\times\reals^n\rightarrow\reals$ is a RCBF on $\mathcal S_H$ in \eqref{eq:inner_safe_set}, then any control law $u(t,x)$ that is locally Lipschitz {\color{black}continuous} in $x$ and piecewise continuous in $t$ such that $u(t,x) \in \boldsymbol\mu_{\textrm{\textnormal{rcbf}}}(t,x), \forall (t,x) \in \mathcal T_H$ will render $\mathcal S_H$ forward invariant. 
\end{lem}
The proof of Lemma~\ref{prior:rcbf_invariance} %is straightforward and 
follows the same steps as \cite[Lem. 4]{garg2020robust}. A version for time-invariant sets is also found in \cite[Thm. 2]{RobustCBFsOld}. We will refer to $u(t,x)\in\boldsymbol\mu_\textrm{rcbf}(t,x)$ as the \textit{RCBF condition}. The consequence of Lemma~\ref{prior:rcbf_invariance} is that if we know 1) $H$ is a RCBF on $\mathcal{S}_H$, 2) the initial condition satisfies 
{\color{black}$x(t_0)\in\mathcal{S}_H(t_0)$},
and 3) 
{\color{black}$\mathcal{S}_H(t)\subseteq\mathcal{S}(t),\forall t\in\mathcal{D}$},
then we immediately know that there exists a safe trajectory beginning at $(t_0,x(t_0))$ satisfying the input constraints. 
{\color{black}We will also extend Lemma~\ref{prior:rcbf_invariance} to the set $\mathcal{S}_H^\textrm{res}$ in certain cases where $\mathcal{S}_H(t) \not\subseteq \mathcal{S}(t)$.}

\color{black}
\begin{rem}
Note that Lemma~\ref{prior:rcbf_invariance} only requires that $H$ is a RCBF on $\mathcal{S}_H$, not on an open subset $\boldsymbol{\mathit\Omega} \supset \mathcal{S}_H$, as is common in the CBF literature. This is intentional, as we seek to develop methods that ensure that the state can never leave $\mathcal{S}_H$. That is, unlike \cite[Prop. 2]{Ames_Journal}, we are not interested in achieving asymptotic stability of $\mathcal{S}_H$ from a larger domain $\boldsymbol{\mathit\Omega}$. %See also \cite[Def. 3]{CBFs_STL} and \cite[Def. 1]{garg2020robust}.
\end{rem}
\color{black}

\section{Control Barrier Functions for Input Constraints} \label{sec:new_cbfs}

Given some constraint function $h$ and associated safe set $\mathcal{S}$, {\color{black}along with input constraints $\mathcal{U}$ and disturbances $\mathcal{W}$,} the goal of this section is to develop a RCBF $H$ and associated 
{\color{black}sets $\mathcal{S}_H$ in \eqref{eq:inner_safe_set} and $\mathcal{S}_H^\textrm{res}$ in \eqref{eq:restricted_safe_set}}
for dynamics relevant to spacecraft, namely relative-degree 2 dynamics. 
The primary challenge addressed by all the subsequent methods is that for constraint functions with high-relative-degree, the input constraints and disturbances must also be incorporated into the form of the RCBF. For example, if $h\in\highdegree^2$, there may exist a state $(t_0,x(t_0))\in\mathcal{S}(t_0)$ such that $h(t_0,x(t_0)) < 0$ and $\dot{h}(t_0,x(t_0)) > 0$. 
Without loss of generality, 
{\color{black}$h$ can be thought of}
as the position of an agent, $\dot{h}$ its velocity, and $\ddot{h}$ its acceleration. This state can be in the zero sublevel set {\color{black}$\mathcal{S}_H$} of some RCBF {\color{black}$H$} only if there exists a control input satisfying the input constraints such that the agent decelerates to $\dot{h}(t,x(t)) = 0$ before leaving $\mathcal{S}$. Thus, a RCBF must be a function of the input constraints, and we seek {\color{black}systematic methods of finding such functions}. 
The following subsections identify three {\color{black}forms of RCBFs}, presented in order of increasing complexity and decreasing conservatism. {\color{black}Each form is developed as a function of $h$ and is applicable under different conditions on the input constraints and disturbances}. Sections~\ref{sec:abs_cbf} and \ref{sec:approach_func} consider only relative-degree $r=2$ constraint functions and require specific system properties, while Section~\ref{sec:predictive_cbf} provides an approach for relative-degree $r\geq 2$ with distinct requirements.

\subsection{Constant Control Authority} \label{sec:abs_cbf}

In this subsection, suppose the constraint function $h$ is of relative-degree $r=2$, and has the special property that $\ddot{h}$ can always be made less than some negative constant (e.g. see \cite[Eq. 17]{CDC21}). {\color{black}Intuitively, this means that the controller can add/remove ``energy'' from the system at a constant rate.} Then, assuming no disturbances, \cite{CDC21} provides the following form for a CBF.

\begin{lem}[{\cite[Example 1]{CDC21}}] \label{lemma:constant_a}
    Suppose $w_{u}\equiv w_{x}\equiv 0$. %$w_{u,\textrm{\textnormal{max}}}=w_{x,\textrm{\textnormal{max}}}=0$. 
    If $h\in \highdegree^2$, $f,g,h$ are time-invariant, and there exists $a_\textrm{\textnormal{max}} > 0$ such that $\forall x\in \mathcal S$,
    \begin{equation}
        \inf_{u\in \mathcal U}\ddot{h}(x,u) \leq -a_{\textrm{\textnormal{max}}} \,{\color{black},} \label{eq:a_max_def}
    \end{equation}
    {\color{black}then} the time-invariant function
    \begin{equation}
        H(x) = \begin{cases} h(x) & \dot{h}(x) < 0 \\ \displaystyle h(x) + \frac{\dot{h}(x)^2}{2a_\textrm{\textnormal{max}}} & \dot{h}(x) \geq 0 \end{cases} \label{eq:constant_a_cbf}
    \end{equation}
    is a CBF on $\mathcal S_H$ in \eqref{eq:inner_safe_set} for the system \eqref{eq:model} {\color{black}for any $\alpha\in\mathcal{K}$}{\color{black}, where $\mathcal{S}_H \equiv \mathcal{S}_H^\textrm{\textnormal{res}}$}. % with $w_u\equiv w_x\equiv 0$.
\end{lem}

Next, we note that \eqref{eq:constant_a_cbf} can be modified to remove the piecewise definition. 
{\color{black}The following} alternate form of \eqref{eq:constant_a_cbf} will be more useful when disturbances are added.

\begin{figure}
    \centering
    \includegraphics[width=0.9\columnwidth,clip,trim={0.25in, 0in, 0.5in, 0.2in}]{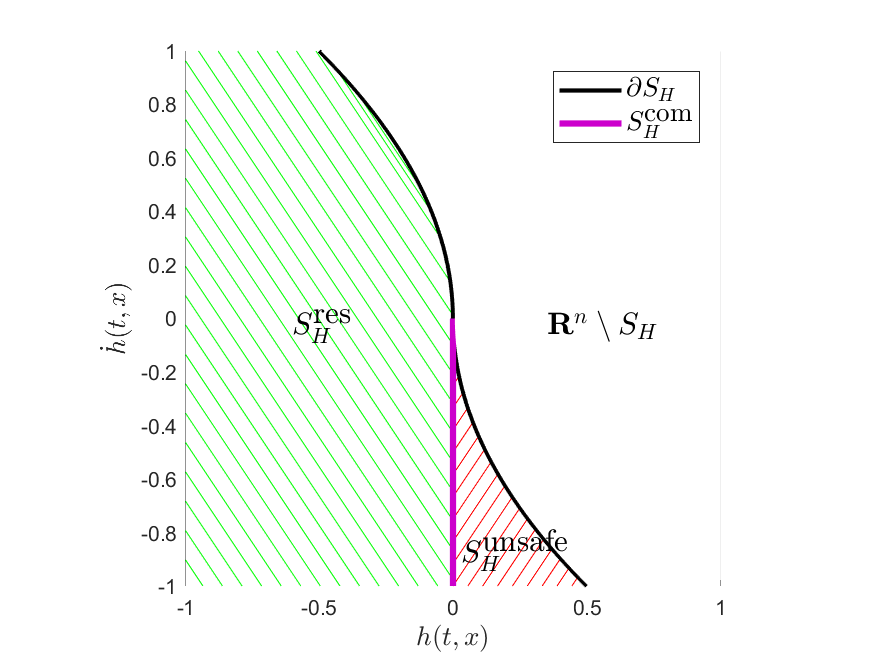}
    \caption{An illustration of the sets $\mathcal{S}_H^\textrm{res}$ and $\mathcal{S}_H^\textrm{unsafe}$ and the manifold $\mathcal{S}_H^\textrm{com}$ between them in magenta. Only states where $h(t,x) \leq 0$ are safe, and only $\mathcal{S}_H^\textrm{res}$ is rendered forward invariant.}
    \label{fig:set_illustration}
\end{figure}

\begin{lem} \label{lemma:absolute_value}
    Suppose that the conditions of Lemma~\ref{lemma:constant_a} hold. Then the function
    \begin{equation}
        H(x) = h(x) + \frac{|\dot{h}(x)|\dot{h}(x)}{2 a_\textrm{\textnormal{max}}} \label{eq:constant_abs_cbf}
    \end{equation}
    is a CBF on $\mathcal{S}_H^\textrm{\textnormal{res}}$ in \eqref{eq:restricted_safe_set} for the system \eqref{eq:model} {\color{black}for any $\alpha\in\mathcal{K}$}. 
    Moreover, any {\color{black}control law $u(t,x)$} that is locally Lipschitz {\color{black}continuous} in $x$ and piecewise continuous in $t$ such that $u(t,x) \in \boldsymbol\mu_\textrm{\textnormal{rcbf}}(t,x), \forall (t,x)\in\mathcal{T}_H^\textrm{\textnormal{res}}$ also renders $\mathcal{S}_H^\textrm{\textnormal{res}}$ forward invariant.
\end{lem}
\begin{pf}
    First note that $H$ in \eqref{eq:constant_abs_cbf} is continuously differentiable {\color{black}(recall that {\color{black}$z:\reals\rightarrow\reals$, $z(\lambda) = \lambda |\lambda|$} is in $\mathcal{C}^1$).} Next, note that the derivative of $H$ is
    \begin{equation}
        \dot{H}(x,u) = \dot{h}(x) + \frac{|\dot{h}(x)|\ddot{h}(x,u)}{ a_\textrm{max}} \,{\color{black}.} \label{eq:time_invariant_dotH}
    \end{equation}
    {\color{black}When $\dot{h}(x) \leq 0$, \eqref{eq:a_max_def} implies that we can always choose a $u$ that renders \eqref{eq:time_invariant_dotH} nonpositive. When $\dot{h}(x) > 0$, \eqref{eq:time_invariant_dotH} reduces to $\dot{H}(x,u) = \dot{h}(x)\left( 1 + \frac{\ddot{h}(x,u)}{a_\textrm{max}} \right)$. Any $u$ such that $\ddot{h}(x,u) \leq -a_\textrm{max}$ will thus render $\dot{H}$ nonpositive, and by assumption in \eqref{eq:a_max_def}, such a $u$ always exists in $\mathcal{U}$ for all $x\in\mathcal{S}$}. Thus, for every $x\in \mathcal{S}_H^\textrm{res}\subseteq \mathcal{S}$, there exists $u(x)\in \mathcal U$ such that $\dot{H}(x,u(x))\leq0\leq \alpha(-H(x))$ for any $\alpha \in \mathcal{K}$, so $H$ satisfies the conditions of a CBF on $\mathcal{S}_H^\textrm{res}$ in Definition~\ref{def:cbf_definition} {\color{black}for any $\alpha\in\mathcal{K}$}.
    
    {\color{black}Next, note that for $H$ as in \eqref{eq:constant_abs_cbf}, the set $\mathcal{S}_H$ in \eqref{eq:inner_safe_set} is not a subset of $\mathcal{S}$ in \eqref{eq:safe_set}; i.e., there exist states $x\in\mathcal{S}_H$ where $H(x) \leq 0$ and $h(x) > 0$, as shown by the red hashed region in Fig.~\ref{fig:set_illustration}. Note that no such states occurred previously for $H$ as in \eqref{eq:constant_a_cbf} due to the piecewise definition.
    Thus, we now divide the set $\mathcal{S}_H$} into two disjoint subsets, the restricted safe set $\mathcal S_H^{\textrm{res}}$ as in \eqref{eq:restricted_safe_set} and the set $\mathcal S_H^{\textrm{unsafe}} \triangleq \{x\in \mathcal S_H \mid h(x) > 0\}$, which is unsafe. Denote the manifold between these two subsets as $\mathcal S_H^{\textrm{com}} \triangleq \{ x\in \mathcal S_H \mid h(x) = 0\}$, illustrated in Fig.~\ref{fig:set_illustration}. It follows from Lemma~\ref{prior:rcbf_invariance} that closed-loop trajectories satisfying $u(t,x)\in\boldsymbol\mu_\textrm{rcbf}(t,x),\forall (t,x)\in\mathcal{T}_H^\textrm{res}$ and the above regularity conditions cannot transition directly from $\mathcal S_H^\textrm{res}$ to $\reals^n\setminus\mathcal{S}_H$ (i.e. cannot cross the black line in Fig.~\ref{fig:set_illustration}). Additionally, a trajectory starting in $\mathcal S_H^{\textrm{res}}$ can only transition directly from $\mathcal S_H^{\textrm{res}}$ to $\mathcal S_H^{\textrm{unsafe}}$ along $\mathcal S_H^{\textrm{com}}$ (i.e. cross the magenta line in Fig.~\ref{fig:set_illustration}).
    \regularversion{Moreover, $\forall x \in \mathcal{S}_H^\textrm{com}$, it holds that $H(x) = |\dot{h}(x)|\dot{h}(x)/(2a_\textrm{max}) \leq 0$, which implies $\dot{h}(x) \leq 0$. 
    Thus, by Nagumo's theorem \cite[Thm 3.1]{Blanchini}, trajectories starting in $\mathcal S_H^{\textrm{res}}$ never directly transition to $\mathcal S_H^{\textrm{unsafe}}$. Since closed-loop trajectories can never transition from $\mathcal{S}_H^\textrm{res}$ to either $\mathcal{S}_H^\textrm{unsafe}$ or $\reals^n\setminus\mathcal{S}_H$, it follows that $\mathcal{S}_H^\textrm{res}$ is rendered forward invariant.}%
    \extendedversion{Divide $\mathcal{S}_H^\textrm{com}$ into two disjoint sets $\mathcal{S}_H^{\textrm{c,1}} = \{x\in\mathcal{S}_H^\textrm{com} \mid H(x) < 0\}$ and $\mathcal{S}_H^{\textrm{c,2}} = \{x\in\mathcal{S}_H^\textrm{com} \mid H(x) = 0\}$. For all $x\in\mathcal{S}_H^{c,1}$, it holds that $\dot{h}(x) < 0$, so trajectories $x(t)$ can never cross the manifold $x\in\mathcal{S}_H^{c,1}$ from inside $\mathcal{S}_H^\textrm{res}$. For all $x\in\mathcal{S}^{c,1}$, it holds that $\dot{h}(x) = 0$. Next, note that any controller satisfying $u(x)\in\boldsymbol\mu_\textrm{rcbf}(x)$ at $x\in\{x\mid H(x) = 0 \textrm{ and } \dot{h}(x) > 0\}$ must yield $\ddot{h}(x) \leq -a_\textrm{max}$. By the above regularity assumptions, it follows that $\ddot{h}(x) \leq -a_\textrm{max} < 0, \forall (x) \in \mathcal{S}_H^{c,2}$, so trajectories $x(t)$ can never cross the manifold $x\in\mathcal{S}_H^{c,2}$ either from inside $\mathcal{S}_H^\textrm{res}$. Thus, trajectories starting in $\mathcal S_H^{\textrm{res}}$ never directly transition across $\mathcal{S}_H^\textrm{com}=\mathcal{S}_H^{c,1}\cap\mathcal{S}_H^{c,2}$ to $\mathcal S_H^{\textrm{unsafe}}$. Since closed-loop trajectories can never transition from $\mathcal{S}_H^\textrm{res}$ to either $\mathcal{S}_H^\textrm{unsafe}$ or $\reals^n\setminus\mathcal{S}_H$, it follows that $\mathcal{S}_H^\textrm{res}$ is rendered forward invariant.}%
    \hfill $\blacksquare$
\end{pf}

\begin{rem}
    If \eqref{eq:a_max_def} holds for all $x \in \mathcal S_H$, then any controller satisfying $u(t,x) \in \boldsymbol\mu_\textrm{\textnormal{rcbf}}(t,x), \forall (t,x)\in\mathcal{T}_H$ also necessarily causes trajectories originating in $\mathcal{S}_H^\textrm{\textnormal{unsafe}}$ to approach $\mathcal S_H^{\textrm{\textnormal{res}}}$ {\color{black}(see also \cite[Rem. 2]{robust_hocbfs})}.
\end{rem}

Note that by definition $\mathcal{S}_H^\textrm{res}$ is a subset of $\mathcal{S}$, so rendering $\mathcal{S}_H^\textrm{res}$ forward invariant also ensures safety for all future time. Moreover, Lemma~\ref{lemma:absolute_value} says that the same CBF condition as in Lemma~\ref{prior:rcbf_invariance} causes $\mathcal{S}_H^\textrm{res}$ to be rendered forward invariant (assuming no disturbances).
In fact, the set $\mathcal{S}_H^\textrm{res}$ for the CBF in \eqref{eq:constant_abs_cbf} is identical to the set $\mathcal{S}_H$ for the CBF in \eqref{eq:constant_a_cbf}, but the form of \eqref{eq:constant_abs_cbf} is mathematically more convenient. 
In particular, the CBF in \eqref{eq:constant_abs_cbf} is of relative-degree 1 everywhere, and captures the rate at which $h$ decreases, unlike the piecewise form in \eqref{eq:constant_a_cbf}.

Next, we consider disturbances. Note that \eqref{eq:constant_abs_cbf} is a function of the constraint function derivative $\dot{h}$, and that in the case of unmatched disturbances (i.e. when $w_x\neq0$), $\dot{h}$ is a function of the disturbance $w_x$ and thus not exactly known. We define the following upper bound on $\dot{h}$:
\begin{align}
    \dot{h}_w(t,x) &\triangleq \max_{||w_x||\leq w_{x,\textrm{max}}} \dot{h}(t,x,w_x)  \label{eq:def_hw} \\ =& \partial_t h(t,x) + \nabla h(t,x) f(t,x) + ||\nabla h(t,x)||w_{x,\textrm{max}} \nonumber
\end{align}
{\color{black}(recall that $h$ is relative-degree 2, so $\nabla h(t,x) g(t,x) \equiv 0$ and thus $\dot{h}$ does not depend on $u, w_u$)} 
and its derivative
\begin{align}
    % deliberately not using \triangleq here
    \ddot{h}_w(t,x,u,&w_u,w_x) = \frac{d}{dt}\dot{h}_w(t,x) \\ &= \partial_t \dot{h}_w(t,x) + \nabla \dot{h}_w(t,x) F(t,x,u,w_u,w_x) \,. \nonumber 
\end{align}
% Note: Since this maximization is just choosing w in the direction matching the appropriate gradient, these maximizations are indeed max, not sup.
% 
Note that $\dot{h}_w$ is a known quantity, while $\ddot{h}_w$ is a function of the unknown quantities $w_u,w_x$ in $F$ {\color{black}in \eqref{eq:model}. Assume $\|\nabla h\|$ does not vanish, so that $\dot{h}_w$ in \eqref{eq:def_hw} is differentiable 
(note that $\|\nabla h\| \equiv 1$ in Section~\ref{sec:spacecraft_functions}).} We are now ready to present the robust formulation of \eqref{eq:constant_a_cbf},\eqref{eq:constant_abs_cbf}.

\begin{thm} \label{thm:constant_noise_cbf}
    Suppose $h\in \highdegree^2$ and there exists $a_\textrm{\textnormal{max}} > 0$ such that $\forall (t,x) \in \mathcal T$, 
    \begin{equation}
        \max_{(w_u,w_x)\in\mathcal{W}} \Big( \inf_{u\in \mathcal U} \ddot{h}_w(t,x,u,w_u,w_x) \Big) \leq - a_\textrm{\textnormal{max}} \,. \label{eq:a_max_noise_def}
    \end{equation}
    Then the function
    \begin{equation}
        H(t,x) = h(t,x) + \frac{|\dot{h}_w(t,x)|\dot{h}_w(t,x)}{2 a_{\textrm{\textnormal{max}}}} \label{eq:constant_noise_cbf}
    \end{equation}
    is a RCBF on $\mathcal S_H^{\textrm{\textnormal{res}}}$ in \eqref{eq:restricted_safe_set} for the system \eqref{eq:model} {\color{black}for any $\alpha\in\mathcal{K}$}. Moreover, any {\color{black}control law $u(t,x)$} that is locally Lipschitz {\color{black}continuous} in $x$ and piecewise continuous in $t$ such that $u(t,x) \in \boldsymbol\mu_\textrm{\textnormal{rcbf}}(t,x), \forall (t,x)\in\mathcal{T}_H^\textrm{\textnormal{res}}$ also renders $\mathcal{S}_H^\textrm{\textnormal{res}}$ forward invariant.
\end{thm}
\begin{pf}
    As in Lemma~\ref{lemma:absolute_value}, $H$ is continuously differentiable, and its derivative satisfies
    \begin{multline}
        \dot{H}(t,\hspace{-0.5pt}x,\hspace{-0.5pt}u,\hspace{-0.5pt}w_u,\hspace{-0.5pt}w_x\hspace{-0.5pt}) \hspace{-2pt}=\hspace{-2pt} \dot{h}(t,\hspace{-0.5pt}x,\hspace{-0.5pt}w_x\hspace{-0.5pt}) + \frac{|\dot{h}_w(t,\hspace{-0.5pt}x)|\ddot{h}_w(t,\hspace{-0.5pt}x,\hspace{-0.5pt}u,\hspace{-0.5pt}w_u,\hspace{-0.5pt}w_x\hspace{-0.5pt})}{a_\textrm{max}} \\  \overset{\eqref{eq:def_hw}}{\leq} \dot{h}_w(t,x) + \frac{|\dot{h}_w(t,x)|\ddot{h}_w(t,x,u,w_u,w_x)}{a_\textrm{max}} \,. \label{eq:constant_noise_cbf_derivative}
    \end{multline}
    By assumption in \eqref{eq:a_max_noise_def}, there exists $u\in\mathcal{U}$ independent of the disturbances $w_u,w_x$ such that {\color{black}$\max_{(w_u,w_x)\in\mathcal{W}}\ddot{h}_w(t,x,u,w_u,w_x) \leq -a_\textrm{max}$. Such a $u$ will render the right hand side of \eqref{eq:constant_noise_cbf_derivative} nonpositive similar to \eqref{eq:time_invariant_dotH} in Lemma~\ref{lemma:absolute_value}, and therefore will also render} $\max_{(w_u,w_x)\in\mathcal{W}} \dot{H}(t,x,u,w_u,w_x)$ nonpositive. {\color{black}Thus,} condition \eqref{eq:rcbf_definition} is satisfied for any $\alpha\in\mathcal{K}$ and all $(t,x)\in\mathcal{T}_H^\textrm{res}$, so $H$ is a RCBF on $\mathcal{S}_H^\textrm{res}$ {\color{black}for any $\alpha\in\mathcal{K}$}. Next, {\color{black}similar to with \eqref{eq:constant_abs_cbf}, for $H$ as in \eqref{eq:constant_noise_cbf}, the set $\mathcal{S}_H(t)$ in \eqref{eq:inner_safe_set} is not a subset of $\mathcal{S}(t)$ in \eqref{eq:safe_set}. As in} Lemma~\ref{lemma:absolute_value}, a trajectory starting in $\mathcal S_H^{\textrm{res}}{\color{black}(t)}$ {\color{black} and satisfying $u(t,x)\in\boldsymbol\mu_\textrm{rcbf}(t,x),\forall (t,x)\in\mathcal{T}_H^\textrm{res}$} can only transition from $\mathcal S_H^{\textrm{res}}{\color{black}(t)}$ to $\mathcal S_H^{\textrm{unsafe}}{\color{black}(t)\triangleq\{x\in\mathcal{S}_H(t)\mid h(t,x) > 0\}}$ along the manifold $\mathcal S_H^{\textrm{com}}{\color{black}(t)\triangleq\{x\in\mathcal{S}_H(t)\mid h(t,x) = 0\}}$ {\color{black}in Fig.~\ref{fig:set_illustration}}. By the same argument as in Lemma~\ref{lemma:absolute_value}, if $x\in \mathcal S_H^{\textrm{com}}{\color{black}(t)}$, then $0 \geq \dot{h}_w(t,x) \geq \dot{h}(t,x,w_x)$, so {\color{black}trajectories starting in $\mathcal{S}_H^\textrm{res}(t)$ cannot leave $\mathcal{S}_H^\textrm{res}(t)$ via $\mathcal{S}_H^\textrm{com}(t)$. Thus, } a controller satisfying $u(t,x)\in\boldsymbol\mu_\textrm{rcbf}(t,x),\forall (t,x)\in\mathcal{T}_H^\textrm{res}$ similarly renders $\mathcal S_H^{\textrm{res}}$ forward invariant. \hfill $\blacksquare$
\end{pf}

Thus, we have presented our first method for rendering trajectories always inside sublevel sets of relative-degree $r=2$ constraint functions under input constraints and disturbances, by constructing a general form of RCBF $H$ in \eqref{eq:constant_noise_cbf} as a function of the constraint function $h$, the input constraints (encoded in $a_\textrm{max}$), and the disturbance bounds.
{\color{black}Note that Theorem~\ref{thm:constant_noise_cbf} showed that \eqref{eq:rcbf_definition} is satisfied for any $\alpha\in\mathcal{K}$, so Section~\ref{sec:switching} will suggest a choice for the free parameter $\alpha\in\mathcal{K}$. The RCBF in Theorem~\ref{thm:constant_noise_cbf}}
is particularly useful for systems similar to the double integrator, as illustrated in \cite{CDC21}. 
It is also easy to check if this method is applicable to a system or not. The largest allowable $a_\textrm{max}$ is 
\begin{align}
    a_{\textrm{max},0} \triangleq - &\max_{(t,x)\in\mathcal{T}} \big[ \partial_t \dot{h}_w(t,x) + \nabla \dot{h}_w(t,x) (f(t,x) \nonumber \\ &\;\;\;\; + g(t,x) u_\textrm{min}(t,x)) + \| \nabla\dot{h}_w(t,x)\| w_{x,\textrm{max}} \nonumber \\ &\;\;\;\; + \|\nabla \dot{h}_w(t,x) g(t,x)\| w_{u,\textrm{max}} \big] \,, \label{eq:a_max0}
\end{align}
where
\begin{equation}
    u_\textrm{min}(t,x) = \argmin_{u\in \mathcal{U}} \nabla \dot{h}_w(t,x) g(t,x) u \,.
\end{equation}
{\color{black}Note that \eqref{eq:a_max0} can be solved offline.} 
If $a_{\textrm{max},0} > 0$, then any $a_\textrm{max} \in (0, a_{\textrm{max},0}]$ satisfies the requirements of Theorem~\ref{thm:constant_noise_cbf}, while if $a_{\textrm{max},0} < 0$, then no such $a_\textrm{max}$ exists for the particular $h$. The form of RCBF in \eqref{eq:constant_noise_cbf} is constructive, since if an $a_\textrm{max} > 0$ exists, then we know that a subset $\mathcal{S}_H^\textrm{res}\subseteq\mathcal{S}$ can be rendered forward invariant, and we have explicit expressions for the restricted safe set $\mathcal{S}_H^\textrm{res}$ in \eqref{eq:restricted_safe_set} and the set of safe control inputs $\boldsymbol\mu_\textrm{rcbf}$ in \eqref{eq:valid_rcbf_control}. 

However, even if an $a_\textrm{max}>0$ does exist, this method can be overly conservative. {\color{black}That is, there may exist $(t,x)\in\mathcal{T}$ such that $(t,x)$ can be robustly rendered inside $\mathcal{S}$ but $(t,x)$ is not in $\mathcal{T}_H^\textrm{res}$ with $H$ as in \eqref{eq:constant_noise_cbf}}, as illustrated in simulation in Section~\ref{sec:spacecraft_functions}. The following section describes an alternative to Theorem~\ref{thm:constant_noise_cbf} that aims at reducing conservatism, and which further treats certain cases where no $a_\textrm{max}$ exists.

\subsection{Variable Control Authority} \label{sec:approach_func}

{\color{black}Using the methodology in the prior section, the set $\mathcal{S}_H^\textrm{res}$ only includes states that can be rendered always inside $\mathcal{S}$ by setting $\ddot{h}_w$ equal to a negative constant $-a_\textrm{max}$.}
For agents with a wide operating range, such as a spacecraft operating at various altitudes, {\color{black}this} restriction may {\color{black}result in an} overly conservative {\color{black}set $\mathcal{S}_H^\textrm{res}$}, or no such $a_\textrm{max}$ may exist. On the other hand, in the spacecraft scenario, {\color{black}the gravity of a central attractive body is known to high precision at every altitude. Similarly, electric/magnetic field strengths, spring force, buoyancy force, and aircraft lift/drag forces are well-characterized across operating ranges. Thus, instead of assuming that $\ddot{h}_w$ is upper bounded by a constant as in \eqref{eq:a_max_noise_def}, suppose it is upper bounded by a known function, denoted $\phi$}. This is the idea central to the following theorem.

\begin{thm} \label{thm:variable_cbf}
    Let $h \in \highdegree^2$ define a safe set as in \eqref{eq:safe_set}. Suppose there exists an invertible {\color{black}and {\color{black}strictly} monotone decreasing} function $\Phi:\reals\rightarrow\reals$, $\Phi \in \continuous^1$, whose derivative is $\Phi'=\phi$ for $\phi:\reals\rightarrow\reals$, such that $\forall (t,x) \in \mathcal T$,
    \begin{equation}
        \max_{(w_u,w_x)\in \mathcal{W}} \inf_{u\in \mathcal U} \ddot{h}_w(t,x,u,w_u,w_x) \leq \phi(h(t,x)) \color{black}<\color{black} 0 \,. \label{eq:phi_requirement}
    \end{equation}
    Let $\Phi^{-1}$ be the function for which $\Phi^{-1}(\Phi(\lambda)) = \lambda, \forall \lambda\in\reals$.
    Then the function 
    \begin{equation}
        H(t,x) = \Phi^{-1}\( \Phi(h(t,x)) - \frac{1}{2}\dot{h}_w(t,x)|\dot{h}_w(t,x)| \) \label{eq:variable_cbf}
    \end{equation}
    is a RCBF on $\mathcal{S}_H^{\textrm{\textnormal{res}}}$ in \eqref{eq:restricted_safe_set} for the system \eqref{eq:model} {\color{black}for any $\alpha\in\mathcal{K}$}. Moreover, any {\color{black}control law $u(t,x)$} that is locally Lipschitz {\color{black}continuous} in $x$ and piecewise continuous in $t$ such that $u(t,x) \in \boldsymbol\mu_\textrm{\textnormal{rcbf}}(t,x), \forall (t,x)\in\mathcal{T}_H^\textrm{\textnormal{res}}$ also renders $\mathcal{S}_H^\textrm{\textnormal{res}}$ forward invariant.
\end{thm}
\begin{pf}
$H$ is a RCBF if $\dot{H}$ satisfies the condition \eqref{eq:rcbf_definition}. 
    By the chain rule, the total derivative of $H$ is
    \color{black}\begin{align} \color{black}
        \dot{H} =&\color{black} (\Phi^{-1})'\big( \Phi(h)-{\textstyle\frac{1}{2}}\dot{h}_w|\dot{h}_w|\big) \( \Phi'(h) \dot{h} - |\dot{h}_w| \ddot{h}_w \) \nonumber \\
        \color{black}\overset{\eqref{eq:variable_cbf}}{=}& \color{black}(\Phi^{-1})'( \Phi(H)) \( \Phi'(h) \dot{h} - |\dot{h}_w|\ddot{h}_w \) \label{eq:var_proof_step1}
    \end{align}\color{black}
    where the arguments of $h, \dot{h}_w, H$ are omitted for brevity.
    By definition, $\Phi'(\cdot) = \phi(\cdot)$, and by the Inverse Function Theorem, $(\Phi^{-1})'(\Phi(\cdot)) = \frac{1}{\phi(\cdot)}$, so {\color{black}\eqref{eq:var_proof_step1} becomes}
    \begin{align*}
        \dot{H} =& \frac{1}{\phi(H)} \hspace{-1pt}\(\hspace{-1pt} \phi(h) \dot{h} - |\dot{h}_w|\ddot{h}_w \hspace{-1pt}\)\hspace{-1pt} 
        \hspace{-1pt}\overset{\eqref{eq:def_hw}}{\leq}\hspace{-2pt} \frac{1}{\phi(H)} \hspace{-1pt}\(\hspace{-1pt} \phi(h) \dot{h}_w - |\dot{h}_w|\ddot{h}_w \hspace{-1pt}\)\hspace{-1pt}.
    \end{align*}
    By assumption in \eqref{eq:phi_requirement}, there exists $u\in \mathcal U$ independent of $w_u,w_x$ such that ${\color{black}\max_{(w_u,w_x)\in\mathcal{W}}}\ddot{h}_w(t,x,u,w_u,w_x) \leq \phi(h(t,x))$. 
    % Also, $\phi(H(t,x))\leq 0$ for all $(t,x)\in\mathcal{T}_H$, so
    Since $\phi({\color{black}H}) \leq 0$, it follows that {\color{black}such a $u$ will render
    $\max_{(w_u,w_x)\in\mathcal{W}}\dot{H}(t,x,u,w_u,w_x)$ with $\dot{H}$} as derived above nonpositive. Since this holds independent of $w_u,w_x$, condition \eqref{eq:rcbf_definition} is satisfied for any $\alpha \in \mathcal{K}$ and all $(t,x)\in\mathcal{T}_H^\textrm{res}$, so $H$ is a RCBF on $\mathcal{S}_H^\textrm{res}$ {\color{black}for any $\alpha\in\mathcal{K}$}.
    
    {\color{black}As in Lemma~\ref{lemma:absolute_value} and Theorem~\ref{thm:constant_noise_cbf}, for $H$ as in \eqref{eq:variable_cbf}, it holds that $\mathcal{S}_H(t)\not\subseteq\mathcal{S}(t)$.} Following the logic of {\color{black}these two theorems}, $\forall (t,x) \in \mathcal T_H^\textrm{res}$, it holds that
    \color{black}\begin{align}\color{black}
        0 \geq&\color{black} \Phi^{-1}\( \Phi(h(t,x)) - \frac{1}{2}\dot{h}_w(t,x)|\dot{h}_w(t,x)| \) \,, \nonumber \\
        \color{black}\Phi(0) \leq&\color{black} \Phi(h(t,x)) - \frac{1}{2}\dot{h}_w(t,x)|\dot{h}_w(t,x)| \,, \label{eq:var_proof_step2}
    \end{align}\color{black}
    since $\Phi$ is {\color{black}monotone decreasing}. % (since $\Phi'=\phi\leq 0$ in \eqref{eq:phi_requirement}). 
    If $x\in\mathcal{S}_H^\textrm{com}(t)$, then $h(t,x)=0$ and {\color{black}\eqref{eq:var_proof_step2}} reduces to $\dot{h}_w(t,x) \leq 0$. Thus, by the same argument as in {\color{black}Theorem~\ref{thm:constant_noise_cbf}}, a controller satisfying $u(t,x)\in\boldsymbol\mu_\textrm{rcbf}(t,x), \forall (t,x)\in\mathcal{T}_H^\textrm{res}$ renders $\mathcal{S}_H^\textrm{res}$ forward invariant. \hfill $\blacksquare$
\end{pf}
% Note to self: the above proof is not straightforward, because the values of w_x and w_u that generate W(x) are not necessarily the same w_x and w_u that maximize \ddot{h}_w. That is, the gradient of H is not necessarily the same direction as the gradient of \dot{h}_w. We could assume w_x, w_u are the values that maximize H in middle section of the proof and redefine \ddot{h}_w to match, but that would be a very convoluted and seemingly senseless definition for \ddot{h}_w. 

\begin{rem}
    In most cases, the function $\phi$ is {\color{black}derived from} the dynamics (e.g. {\color{black}$\phi$ might represent} a potential force {\color{black}that can be read directly from the equations of motion}), in which case $\Phi$ can be any anti-derivative of $\phi$. The results are invariant under different constants of integration. For instance, gravity may be described by $\phi(\lambda) = -\frac{\mu}{\lambda^2}$, in which case either $\Phi(\lambda) = \frac{\mu}{\lambda}$ or $\Phi(\lambda) = \frac{\mu}{\lambda} - \frac{\mu}{\lambda_0}$ for fixed $\lambda_0$ meets the requirements of Theorem~\ref{thm:variable_cbf}.
\end{rem}

Thus, we have presented our second form of RCBF that is applicable to relative-degree $r=2$. {\color{black}Intuitively, the function $\Phi$ in \eqref{eq:phi_requirement} is usually a potential field, in which states below a certain potential value are unsafe. One can think of the argument of $\Phi^{-1}$ in \eqref{eq:variable_cbf} as analogous to the sum of potential energy $-\Phi(h)$ and kinetic energy $\frac{1}{2}\dot{h}_w^2$; the inverse of this quantity provides an ``effective constraint value'' $H$ in the same units as the original constraint $h$. Using this analogy, Theorem~\ref{thm:variable_cbf} gives a condition for ensuring an agent moving in this potential field never falls below the minimum safe potential threshold. If an expression for potential energy is known, then finding $\Phi$ is often straightforward, as shown in Section~\ref{sec:spacecraft_functions}, but this may be difficult otherwise}. 
There is no general formula for $\Phi$ as there was for $a_\textrm{max}$ in \eqref{eq:a_max0}{\color{black}, though $\Phi$ can possibly be learned \cite{jin2020neural}}. 

{\color{black}We note that assumptions \eqref{eq:a_max_noise_def} and \eqref{eq:phi_requirement} may appear to be restrictive assumptions, but are reasonable for many systems. {\color{black}Similar} assumptions are implicit in \cite[Eqs. 11-14]{Recursive_CBF} and \cite[Eq. 16]{robust_hocbfs}{\color{black}, except that the assumptions in \cite{Recursive_CBF,robust_hocbfs} only apply to $\mathcal{S}_H^\textrm{res}$ rather than to $\mathcal{S}$, and are relaxed in $\interior(\mathcal{S}_H^\textrm{res})$ by using one additional class-$\mathcal{K}$ function}. 
In fact, $H$ as in \eqref{eq:variable_cbf} can be expressed using the conventions of \cite{Recursive_CBF,robust_hocbfs} as $\psi_1(t,x) = \dot{h}_w(t,x) - \sqrt{2 (\Phi(h(t,x))-\Phi(0))}$. However, in this case, $\psi_1$ no longer has an interpretation as energy, and this form is problematic for the robustness strategy in \cite{robust_hocbfs} because the square-root function is not differentiable at the origin.}

\subsection{General Case Control Authority} \label{sec:predictive_cbf}

We now consider the case where a function $\Phi$ may not exist, or may still yield overly conservative results. To accomplish this, we suppose there is a known control law $u^*$, called the \textit{nominal evading maneuver} in \cite{Predictive_CBF}, which encourages safety by driving the agent towards the interior of $\mathcal{S}$. {\color{black}For instance, $u^*$ can be a controller which drives an agent away from an obstacle, but which does not necessarily provide for stability or convergence to an objective, such as the controller in \eqref{eq:sample_control_law}.} The core idea of this subsection is to propagate the state forward in time using the nominal evading maneuver, and analyze the resultant trajectory for safety. If the propagated trajectory from $(t_0,x(t_0))$ is always in $\mathcal{S}$, then we conclude that $(t_0,x(t_0))$ should be in the inner safe set. This is formalized mathematically for the case without disturbances in the following lemma, adapted from \cite{CDC21} (see also \cite[Thm. 2]{Predictive_CBF}, though that is more restrictive).

\begin{lem}[{\cite[Thm. 1]{CDC21}}] \label{prior:predictive_cbf_thm}
    Suppose $w_x\equiv w_u\equiv 0$. Suppose $u^*:\reals^n\rightarrow \mathcal U$ is locally Lipschitz continuous and $f,g,h$ are time-invariant. Suppose the function $H:\reals^n\rightarrow\reals$, defined as follows, exists $\forall x \in \mathcal S$,
    \begin{equation}
        H(x) = \max_{\beta\geq 0} h(y(\beta)) \label{eq:H_predictive}
    \end{equation}
    where $y(\beta)$ is the solution {\color{black}at time $\beta$} to
    \begin{equation}
        \dot{y} = f(y) + g(y)u^*(y), \;\; y(0) = x \,.
    \end{equation}
    Then $H$ is a CBF {\color{black}in the sense of \cite[Def. 2.3]{Additive_CBFs}} on $\mathcal{S}_H$ in \eqref{eq:inner_safe_set} for the system \eqref{eq:model} {\color{black}for any $\alpha\in\mathcal{K}$}{\color{black}, where $\mathcal{S}_H\equiv\mathcal{S}_H^\textrm{\textnormal{res}}$}.
\end{lem}

Note that \cite[Thm. 1]{CDC21} requires an extended notion of CBF {\color{black}as in \cite{Additive_CBFs} because \eqref{eq:H_predictive} may not be continuously differentiable; this extended definition} is not used subsequently {\color{black}to avoid confusion in Section~\ref{sec:switching}}. 
The extension of Lemma~\ref{prior:predictive_cbf_thm} to the time-varying case is straightforward, but accounting for the presence of disturbances introduces several more restrictions, as we will show. 

For clarity, in this section, we will use the notation $\frac{d h}{d t}, \frac{\partial h}{\partial t}, \frac{\partial h}{\partial x}$ instead of $\dot{h},\partial_t h, \nabla h$, respectively. Suppose existence of a {\color{black}control law (the proposed nominal evading maneuver)} $u^*:\mathcal{D}\times\reals^n\rightarrow\reals^m$ that is locally Lipschitz continuous in $x$ and piecewise continuous in $t$. 
We {\color{black}define}
the function $\chi:\reals_{\geq 0}\times\mathcal{D}\times\reals^n\rightarrow \reals^n$ %, defined
as the solution to the following initial value problem
\begin{align}
    \chi(\beta, t_0, x&_0) = y(\beta), \, \textrm{where }  y(0) = x_0,  \Theta(0) = I,  \theta(0) = 0, \nonumber \\
    Y(\beta, t_0,& y) \triangleq f(t_0+\beta,y) + g(t_0+\beta,y)u^*(t_0+\beta,y) \,, \nonumber \\ 
    \frac{dy}{d\beta} &= Y(\beta, t_0, y) \label{eq:ode_with_noise} \,, \\
    \frac{d\theta}{d\beta} &= \frac{\partial Y(\beta, t_0, y)}{\partial t_0} \,, \nonumber \\
    \frac{d\Theta}{d\beta} &= \frac{\partial Y(\beta, t_0, y)}{\partial y} \Theta \nonumber \,.
\end{align}

Here, $(t_0,x_0)$ is some given state, $\beta$ is the amount of time in the future from $t_0$ we wish to propagate the trajectory, and $\chi$ is the propagated state. Since $\beta$ is ``time since $t_0$'', we use the absolute time $t_0+\beta$ in the arguments of $f,g,u^*$ in \eqref{eq:ode_with_noise}. The solution to \eqref{eq:ode_with_noise} also gives us $\theta$ and $\Theta$, which we will show are the sensitivities (i.e. derivatives) of $\chi(\beta,t_0,x_0)$ with respect to $t_0$ and $x_0$, respectively. Let $\theta(\beta{\color{black},t_0,x_0})$ and $\Theta(\beta{\color{black},t_0,x_0})$ denote the values of $\theta$ and $\Theta$, respectively, at $\beta$.
To determine safety, we are interested in the values of $h(t_0+\beta, \chi(\beta,t_0,x(t_0)))$ for various times in the future $\beta$.

First, we need to make some remarks about the derivatives of $\chi$ from \eqref{eq:ode_with_noise} in the following lemma.
\begin{lem} \label{lemma:derivatives_of_chi}
    Suppose $f,g,u^*\in\continuous^1$, and $\chi$ in \eqref{eq:ode_with_noise} exists everywhere in a neighborhood of $\beta, t_0, x_0$. Then the partial derivatives of $\chi$ are \allowdisplaybreaks
    \begin{align} 
        \frac{\partial \chi(\beta, t_0, x_0)}{\partial \beta} &= Y(\beta, t_0, \chi(\beta, t_0, x_0)) \,, \label{eq:Hp_partial_beta} \\
        \frac{\partial \chi(\beta, t_0, x_0)}{\partial t_0} &= \theta(\beta{\color{black},t_0,x_0}) \,, \label{eq:Hp_partial_t0} \\
        \frac{\partial \chi(\beta, t_0, x_0)}{\partial x_0} &= \Theta(\beta{\color{black},t_0,x_0}) \,. \label{eq:Hp_partial_x0}
    \end{align}
\end{lem}
\extendedversion{
\begin{pf}
    First, the partial derivative with respect to $\beta$ in \eqref{eq:Hp_partial_beta} follows immediately from \eqref{eq:ode_with_noise}, since $y$ in \eqref{eq:ode_with_noise} evolves with respect to $\beta$. 
    
    Second, note that at $\beta=0$, $\chi(0,t_0,x_0) = x_0$, so $\frac{\partial}{\partial t_0}[\chi(0,t_0,x_0)] = 0$. Also, since $f,g,u^*$ are continuously differentiable, we can describe the evolution of the partial derivative with $\beta$ as follows
    \begin{align*}
        \frac{d}{d\beta}\[\frac{\partial}{\partial t_0} [\chi] \] &= \frac{\partial}{\partial t_0} \[ \frac{d}{d\beta}[\chi] \] = \frac{\partial}{\partial t_0}\[ Y\] 
    \end{align*}
    where we omit the arguments for brevity. Thus, $\frac{\partial}{\partial t_0}[\chi]$ in \eqref{eq:Hp_partial_t0} is given exactly by the construction of $\theta$ in \eqref{eq:ode_with_noise}.
    
    Third, note that since $\chi(0,t_0,x_0) = x_0$, it follows that $\frac{\partial}{\partial x_0}[\chi(0,t_0,x_0)] = I$. We can describe the evolution of the partial derivative with $\beta$ similarly to the prior case as
    \begin{align*}
        \frac{d}{d\beta}\[\frac{\partial}{\partial x_0} [\chi] \] &= \frac{\partial}{\partial x_0} \[ \frac{d}{d\beta}[\chi] \] = \frac{\partial}{\partial x_0}\[ Y\] = \frac{\partial Y}{\partial y} \underbrace{\frac{\partial y}{\partial x_0}}_{=\Theta} \,.
    \end{align*}
    Noting that $\chi=y$, the above equation is a linear ODE in $\beta$ for $\frac{\partial y}{\partial x_0}$, and this ODE takes the exact same form as the equation for $\Theta$ in \eqref{eq:ode_with_noise}. Thus, $\frac{\partial}{\partial x_0}[\chi]$ in \eqref{eq:Hp_partial_x0} is also given by $\Theta$ in \eqref{eq:ode_with_noise}. This completes the proof. \hfill $\blacksquare$
\end{pf}
}

\regularversion{The proof of Lemma~\ref{lemma:derivatives_of_chi} is 
{\color{black}found in \cite[Lemma~\ref{lemma:derivatives_of_chi}]{extended}}.}
A similar result is also given in \cite[Thm 6.1]{ode_book} {\color{black}and \cite[Eq. 23]{backup_controller}}. Note that whereas related work in \cite{Predictive_CBF} solved for $\chi$ explicitly, we assume that in most cases no explicit expression will exist, so the ODEs for $\theta$ and $\Theta$ will be necessary for implementation.
Next, define the set 
\begin{equation} \label{eq:noisy_u_set}
    \mathcal U_w = \left\lbrace u\in \reals^m \,\Big|\, \textstyle \frac{||u|| + \lambda w_{u,\textrm{max}}}{||u||} u \in \mathcal U, \forall \lambda \in [-1, 1] \right\rbrace , %\subset \mathcal U
\end{equation}
which represents a subset of the allowable control inputs with a margin for the disturbance. 

Next, note that $H$ in \eqref{eq:H_predictive} was defined as a maximization. Proving Lemma~\ref{prior:predictive_cbf_thm} and its robust extension requires us to find the maximizer times. Thus, given a $u^*$, define the set-valued function $\mathcal{B}:\mathcal{D}\times\reals^n\rightarrow\reals$ as
\begin{equation}
    \mathcal{B}(t,x) \triangleq \big\lbrace \argmax_{\beta\geq0} h(t+\beta, \chi(\beta, t, x)) \big \rbrace \,. \label{eq:maximizers}
\end{equation}
    % Important: This must be a \max, not a \sup
Note that $\mathcal{B}$ could have finitely many elements or could be dense. If no maximizer exists along the trajectory $\chi$ starting from $(t,x)$ (e.g. $h(t+\beta,\chi(\beta,t,x))$ is unbounded or asymptotically approaches its supremum), then the set $\mathcal{B}(t,x)$ is empty.
{\color{black}We now present some assumptions and notations, followed by the robust extension of \cite[Rem. 1]{CDC21}.}

\color{black}
\begin{assum} \label{as:standard}
For the remainder of this subsection, assume \color{black} that $\chi\in\continuous^1$, {\color{black}$w_x \equiv 0$}, $h\in\highdegree^r$ for $r\geq 2$, $\mathcal U_w$ is nonempty, and $u^*:\mathcal D\times \reals^n\rightarrow \mathcal U_w$. {\color{black}Assume also} that for all $(t,x)\in\mathcal{T}$ the set $\mathcal{B}(t,x)$ in \eqref{eq:maximizers} is nonempty and contains at most one nonzero element. {\color{black}Denote the nonzero element of $\mathcal{B}(t,x)$, if it exists, as $\beta^*(t,x)$.} 

Let $\frac{\partial h(\cdot)}{\partial \lambda_t}, \frac{\partial h(\cdot)}{\partial \lambda_x}$ refer to the partial derivative of $h$ with respect to its first and second arguments, respectively, evaluated at $(\cdot)$. This is to remove ambiguity when $(\cdot)$ includes functions of $t$ and $x$. Similarly, define $\frac{\partial \chi(\cdot)}{\partial \lambda_\beta},\frac{\partial \chi(\cdot)}{\partial \lambda_{t_0}},\frac{\partial \chi(\cdot)}{\partial \lambda_{x_0}}$. For compactness, we also abbreviate certain function arguments as $a_1$, $a_2$, $a_3$, defined by underbraces in the following equations. Finally, note that $w_x$ is omitted from the arguments of $\dot{h}$, {\color{black}$\dot{H}$,} and $F$ {\color{black}from} \eqref{eq:model} in {\color{black}the following equations}, because we require $w_x\equiv 0$.
\end{assum}

\color{black}
\begin{lem} 
    \label{lemma:Hpred_derivative_exact}
    Let Assumption~\ref{as:standard} hold. Consider the 
    function $H:\mathcal{D}\times\reals^n\rightarrow\reals$ defined as
    \begin{equation}
        H(t,x) = \max_{\beta \geq 0} h(t+\beta, \chi(\beta,t,x)) \,. \label{eq:predictive_rcbf}
    \end{equation}
    Suppose the control law $u(t,x)$ is Lebesgue-integrable, and let $x(t)$ be a trajectory of \eqref{eq:model} under $u(t,x)$. Then $H(t,x(t))$ is absolutely continuous in $t$ and satisfies
    \begin{multline}
        \dot{H}(t,x,u,w_u) = \hspace{-4pt} \max_{\beta_c \in \mathcal{B}(t,x)} \[ \frac{\partial h(a_1)}{\partial \lambda_t} + \frac{\partial h(a_1)}{\partial \lambda_x} \theta(\beta_c,t,x) \right. \\ \left. + \frac{\partial h(a_1)}{\partial \lambda_x} \Theta(\beta_c,t,x) F(t,x,u,w_u) \] \label{eq:etadot_pred}
    \end{multline}
    almost everywhere along $x(t)$, where $\theta,\Theta$ are as in \eqref{eq:ode_with_noise} {\color{black}and $a_1$ is as in \eqref{eq:Hequiv}}.
\end{lem}
\begin{pf} \allowdisplaybreaks
    {\color{black}First, since $\dot{x}$ satisfies \eqref{eq:model}, any trajectory $x(t)$ must be absolutely continuous \cite[Eq. C.2]{Sontag}. Since $\mathcal{B}(t,x)$ is assumed nonempty, $H$ in \eqref{eq:predictive_rcbf} always exists, and on any compact interval $[t_1,t_2]$ the elements $\beta_c$ of $\mathcal{B}(t,x(t))$ are bounded. Note that $H$ in \eqref{eq:predictive_rcbf} is equivalent to 
    \begin{equation}
        H(t,x) \equiv h(\underbrace{t+\beta_c, \chi(\beta_c,t,x)}_{\color{black}=a_1}) \label{eq:Hequiv}
    \end{equation}
    for any $\beta_c \in \mathcal{B}(t,x)$. Since $h$ and $\chi$ are both continuously differentiable in all arguments, $\beta_c$ is locally bounded, and $x(t)$ is absolutely continuous, it follows that $H(t,x(t))$ in \eqref{eq:Hequiv}, and therefore \eqref{eq:predictive_rcbf} also, is absolutely continuous in $t$.} 
    
\begin{figure}
    \centering
    \includegraphics[width=\columnwidth]{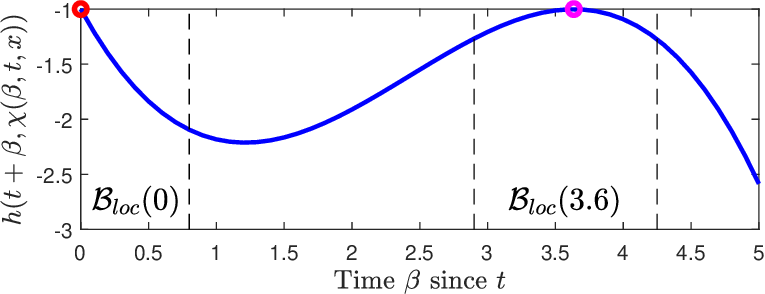}
    \caption{An example trajectory of $h(t+\beta,\chi(\beta,t,x))$ with two maximizers, $\{0, 3.6\}$. Each maximizer $\beta_c$ is surrounded by a set $\mathcal{B}_{loc}(\beta_c)$ on which $\beta_c$ is the unique maximizer. At time $t+\Delta t$ for small $\Delta t$, the trajectory may only have a single global maximizer, but both sets $\mathcal{B}_{loc}$ will still each contain a local maximizer, which is used to compute $\frac{d\beta_c}{dt}$ for each $\beta_c$.}
    \label{fig:predictive}
\end{figure}
    
    \color{black}
    {\color{black}Next, to prove \eqref{eq:etadot_pred}, we are interested in how each maximizer $\beta_c \in \mathcal{B}(t,x)$ varies with $(t,x)$, and then we will study how the value of $H$ in \eqref{eq:Hequiv} varies with $\beta_c$.}
    Since $\mathcal{B}$ is assumed to contain at most one nonzero element, $\mathcal{B}$ cannot be dense. Thus, there exist open subsets of $\reals_{\geq 0}$, denoted $\mathcal{B}_{loc}(\beta_c)${\color{black},} containing only one element $\beta_c$ of $\mathcal{B}(t,x)$. These sets are visualized in Fig.~\ref{fig:predictive}. Within their respective sets $\mathcal{B}_{loc}(\beta_c)$, each $\beta_c\in\mathcal{B}(t,x)$ is a strict maximizer. {\color{black}Thus,} given a $\beta_c\in\mathcal{B}$, the derivative of $\beta_c(t,x)$ with respect to some variable is equivalent to the derivative of the maximizer of \eqref{eq:predictive_rcbf} restricted to $\beta\in\mathcal{B}_{loc}(\beta_c)$ with respect to that variable. {\color{black}We analyze these derivatives in {\color{black}three} cases.} First, if $\beta_c=0\in\mathcal{B}(t,x)$, then $\beta_c$ is a strict maximizer on $\mathcal{B}_{loc}(0) = [0,c)$ for some $c$, and it must hold that {\color{black}$\dot{h}(t+\beta_c,\chi(\beta_c,t,x))\big|_{\beta_c=0} \leq 0$. Note also the equivalency $\color{black}\dot{h}(t+\beta_c,\chi(\beta_c,t,x))\big|_{\beta_c=0} \equiv \frac{dh(a_1)}{d\beta_c}\big|_{\beta_c=0} \equiv \dot{h}(t,x)$, and note that this quantity is independent of $u$, as $h \in \highdegree^r$ for $r\geq 2$ (note how unlike \cite{CDC21}, we do not allow $h\in\highdegree^1$ here). For the first case, further suppose that $\dot{h}(t,x) < 0$ strictly. Then} at an infinitesimal time in the future $t+\Delta t$ {\color{black}(or the past if $\Delta t < 0$)}, the point $\beta_c = 0$ will still be a strict maximizer on $\mathcal{B}_{loc}(0)$, so it must hold that $\frac{d \beta_c}{d t} {\color{black}\big|_{\beta_c=0}} \color{black}= \lim_{\Delta t \rightarrow 0} \frac{\beta_c(t+\Delta t, x(t+\Delta t)) - \beta_c(t,x(t))}{\Delta t} = \lim_{\Delta t\rightarrow 0}\frac{0 - 0}{\Delta t}\color{black} = 0$. {\color{black}Second, if $\beta_c = 0$ and $\dot{h}(t,x) = 0$, then it holds equivalently that $\frac{dh(a_1)}{d\beta_c}\big|_{\beta_c = 0} \equiv \dot{h}(t,x) = 0$. Third}, if $\beta_c \neq 0,\beta_c\in\mathcal{B}(t,x)$, then $\beta_c\color{black}=\beta^*(t,x)\color{black}$ is a maximizer of the $\continuous^1$ function $h$ on an open interval $\mathcal{B}_{loc}(\beta_c) = (c_1,c_2)$ for some $0<c_1<c_2$, so it must hold that the derivative of $h$ at the maximizer is zero, i.e. $\frac{d h(a_1)}{d \beta_c} {\color{black}\big|_{\beta_c=\beta^*}} = 0$. It follows that $\frac{d h(a_1)}{d \beta_c}\frac{d \beta_c}{d t} = 0$ in {\color{black}all three} cases and regardless of $u$ and $\beta_c$, which is the result we will need {\color{black}in \eqref{step5}} below. 
    Next, note that $\dot{H}$ satisfies
    \begin{align}
        \dot{H}(t&,x{\color{black},u,w_u}) = \frac{d H(t,x)}{dt} = \frac{d h(a_1)}{d t} \nonumber \\
        =& \frac{\partial h(a_1)}{\partial \lambda_t}\frac{d (t+\beta_c)}{dt} + \frac{\partial h(a_1)}{\partial \lambda_x}\frac{d \chi(\beta_c, t, x)}{d t} \nonumber  \\
        =& \frac{\partial h(a_1)}{\partial \lambda_t}\(1+\frac{d\beta_c}{dt}\) + \frac{\partial h(a_1)}{\partial \lambda_x} \( \frac{\partial \chi(\beta_c, t, x)}{\partial \lambda_\beta}\frac{d\beta_c}{dt}  \nonumber \right. \\ & \left. \;\; + \frac{\partial \chi(\beta_c, t, x)}{\partial \lambda_{t_0}} + \frac{\partial \chi(\beta_c, t, x)}{\partial \lambda_{x_0}}\frac{dx}{dt} \) \nonumber \\ %\label{eq:new_step}
    % \end{align}
    % for one of the $\beta_c\in\mathcal{B}(t,x)$ \cite[Sec. II]{Additive_CBFs}. 
    % This simplifies to
    % \begin{align}
    %     \dot{H}(&{\color{black}\,\cdot\,}) 
        \overset{\eqref{eq:model}}{=}& \(\frac{\partial h(a_1)}{\partial \lambda_t} + \frac{\partial h(a_1)}{\partial \lambda_x} \frac{\partial \chi(\beta_c, t, x)}{\partial \lambda_\beta} \) \frac{d\beta_c}{dt} + \frac{\partial h(a_1)}{\partial \lambda_t} \nonumber \\ &\hspace{-5pt} +\hspace{-1pt} \frac{\partial h(a_1)}{\partial \lambda_x}\frac{\partial \chi(\beta_c, t, x)}{\partial \lambda_{t_0}} \hspace{-2pt}+\hspace{-2pt} \frac{\partial h(a_1)}{\partial \lambda_x}\frac{\partial \chi(\beta_c, t, x)}{\partial \lambda_{x_0}}F(t,x,u,w_u) \nonumber \\
        =\,& \underbrace{\frac{dh(a_1)}{d\beta_c} \frac{d\beta_c}{dt}}_{=0} + \frac{\partial h(a_1)}{\partial \lambda_t} + \frac{\partial h(a_1)}{\partial \lambda_x}\underbrace{\frac{\partial \chi(\beta_c, t, x)}{\partial \lambda_{t_0}}}_{\color{black}=\theta(\beta_c,t,x)} \nonumber \\ &\;\; + \frac{\partial h(a_1)}{\partial \lambda_x}\underbrace{\frac{\partial \chi(\beta_c, t, x)}{\partial \lambda_{x_0}}}_{\color{black}=\Theta(\beta_c,t,x)} F(t,x,u,w_u) \label{step5}
    \end{align}
    for one of the $\beta_c\in\mathcal{B}(t,x)$ \cite[Sec. II]{Additive_CBFs}. {\color{black}Specifically, if $\mathcal{B}(t,x(t))$ contains $N$ elements at time $t$, then there are $N$ sets $\mathcal{B}_{loc}$. At an infinitesimal time in the future $t+\Delta t$, each set $\mathcal{B}_{loc}$ will still contain a local maximizer, and at least one of these will still be a global maximizer (for sufficiently small $\Delta t$). The derivative $\dot{H}$ corresponds to the rate of change of whichever local maximizer(s) is the global maximizer at both $t$ and $t+\Delta t$, i.e. the $\beta_c \in \mathcal{B}(t,x)$ that maximizes \eqref{step5}. This is then encoded in \eqref{eq:etadot_pred}.} \hfill $\blacksquare$
\end{pf}

{\color{black}Now that we have an expression for the total derivative \eqref{eq:etadot_pred} of $H$ in \eqref{eq:predictive_rcbf}, the next lemma derives an upper bound on this derivative that we then use in Theorem~\ref{thm:predictive_cbf} to prove that the corresponding set $\mathcal{S}_H$ can be rendered forward invariant while satisfying the input constraints.}

\color{black}
\begin{lem} 
    \label{lemma:Hpred_derivative_inequality}
    Suppose the assumptions of Lemma~\ref{lemma:Hpred_derivative_exact} hold. Then $H$ in \eqref{eq:predictive_rcbf} satisfies
    \begin{equation}
        \dot{H}(t,x,u,w_u) \leq \begin{cases} 
            0 \hspace{0.87in} \mathcal{B}(t,x(t)) = \{ 0 \} \\ 
            \begin{aligned} \max\{& 0,\, q(\beta^*(t,x),t,x) [ u + w_u \\  - u&^*(t,x) ] \} \;\;\;\; \mathcal{B}(t,x(t)) \neq \{ 0\}\end{aligned} 
            \end{cases}  \hspace{-16pt} \label{eq:etadot_simplified}
    \end{equation}
    almost everywhere along $x(t)$, where
    \begin{equation}
        q(\beta,t,x) \triangleq \nabla h(t+\beta,\chi(\beta,t,x)) \Theta(\beta,t,x) g(t,x) \,. \label{eq:def_q}
    \end{equation}
\end{lem}

\color{black}
\begin{pf}
    Let $\beta_c$ be any element of $\mathcal{B}(t,x)$. 
    Since $\beta_c$ is a maximizer of \eqref{eq:predictive_rcbf}, it must hold {\color{black}\cite[Eq. 11.35]{optimization}} that 
    \begin{align}
        0 \geq& \frac{d}{d \beta_c}[h(\underbrace{t+\beta_c, \chi(\beta_c, t, x)}_{=a_1})] \nonumber \\ =& \frac{\partial h(a_1)}{\partial \lambda_t} + \frac{\partial h(a_1)}{\partial \lambda_x}\frac{\partial \chi(\beta_c, t, x)}{\partial \lambda_\beta}
        \label{step1} \,.
    \end{align}
    Next, we define the quantity $\kappa$ as follows,
    \begin{align*}
        \kappa &\triangleq h(t + \beta_c, \chi(\beta_c, t, x))
    \end{align*}
    {\color{black}and note that $\kappa$ has the equivalent form}
    \begin{align*}
        \kappa &\equiv h(\underbrace{t + \beta_c, \chi(\underbrace{\beta_c-\tau, t+\tau, \chi(\tau, t, x)}_{=a_2})}_{=a_3})
    \end{align*}
    for any $\tau \in [0, \beta_c]$. Since $\kappa$ is constant with respect to $\tau$, it follows that $\frac{d}{d\tau}(\kappa)=0$, so
    \begin{equation*}
        % \frac{d \kappa}{d \tau}
        0= \frac{\partial h(a_3)}{\partial \lambda_x} \( -\frac{\partial \chi(a_2)}{\partial \lambda_\beta} + \frac{\partial \chi(a_2)}{\partial \lambda_{t_0}} + \frac{\partial \chi(a_2)}{\partial \lambda_{x_0}} \frac{\partial \chi(\tau, t, x)}{\partial \lambda_\beta}\) .
    \end{equation*}
    At $\tau = 0$, this becomes
    \begin{align}
        0&=\frac{\partial h(a_1)}{\partial \lambda_x}\( - \frac{\partial\chi(\beta_c, t, x)}{\partial \lambda_\beta} + \frac{\partial\chi(\beta_c,t,x)}{\partial \lambda_{t_0}} \right.\nonumber \\ &\;\;\;\;\left. + \frac{\partial \chi(\beta_c,t,x)}{\partial \lambda_{x_0}} Y(0,t,x) \) \nonumber \\
        &\overset{\eqref{step1}}{\geq} \frac{\partial h(a_1)}{\partial \lambda_t} + \frac{\partial h(a_1)}{\partial \lambda_x} \(\frac{\partial \chi(\beta_c, t, x)}{\partial \lambda_{t_0}} \right. \nonumber \\ &\;\;\;\;\left. + \frac{\partial \chi(\beta_c, t, x)}{\partial \lambda_{x_0}} Y(0, t, x_0) \) .
        \label{step4}
    \end{align}
    
    {\color{black}$\dot{H}$ in \eqref{step5} then} simplifies to
    \begin{align}
        \dot{H}(\,\cdot\,)
        % \overset{\substack{\eqref{step5} \\ \eqref{step4}}}{\leq} 
        \overset{\eqref{step4}}{\leq} 
        &\frac{\partial h(a_1)}{\partial \lambda_x}\frac{\partial \chi(\beta_c, t, x)}{\partial \lambda_{x_0}}\(F(t,x,u,w_u) - Y(0,t,x) \) \nonumber \\
        \overset{{\color{black}\eqref{eq:model},\eqref{eq:ode_with_noise}}}{=}\hspace{-6.5pt}& \hspace{6.5pt} \frac{\partial h(a_1)}{\partial \lambda_x}\frac{\partial \chi(\beta_c, t, x)}{\partial \lambda_{x_0}}g(t,x)\(u + w_u - u^*(t,x) \) \nonumber \\
        \color{black}\overset{\eqref{eq:def_q}}{=}& \color{black}q(\beta_c,t,x)(u + w_u - u^*(t,x)) \,. \label{eq:end_of_proof}
    \end{align}
    {\color{black}As in Lemma~\ref{lemma:Hpred_derivative_exact}, $\dot{H}$ is upper bounded by the maximum of \eqref{eq:end_of_proof} over all $\beta_c \in \mathcal{B}(t,x)$. Next, recall that $\Theta(0,t,x) = I$ in \eqref{eq:ode_with_noise}, so $q(0,t,x) = \nabla h(t,x) g(t,x)$ in \eqref{eq:def_q}. By definition of $h$ being of relative-degree $r \geq 2$, it follows that $q(0,t,x) = 0$ for any $(t,x)$. Therefore, \eqref{eq:end_of_proof} is at most 0 when $\beta_c = 0$. It follows that $\dot{H}$ in \eqref{eq:etadot_pred} is at most 0 when $\beta_c = 0$ is the only element of $\mathcal{B}$, and $\dot{H}$ is upper bounded by the second case of \eqref{eq:etadot_simplified} when there is a nonzero $\beta_c = \beta^* \in \mathcal{B}$.} \hfill $\blacksquare$
\end{pf}

\color{black}
That is, due to the $\max$ function, $H$ in \eqref{eq:predictive_rcbf} is potentially non-smooth, so the partial derivatives $\partial_t H$ and $\nabla H$ are not well defined. Instead, Lemmas~\ref{lemma:Hpred_derivative_exact}-\ref{lemma:Hpred_derivative_inequality} provide expressions for the total derivative $\dot{H}$.
However, \eqref{eq:etadot_simplified} still contains the unknown disturbance $w_u$, so based on \eqref{eq:etadot_simplified} and with a slight abuse of notation, we now re-define
\begin{equation}
    W(t,x) = \| q(\beta^*(t,x),t,x) \| w_{u,\textrm{max}} \,, \label{eq:new_W}
\end{equation}% %\vspace{-10pt}
\begin{align}
    \boldsymbol\mu&_\textrm{rcbf}(t,x) = \{ u \in \mathcal{U} \mid  \textrm{if } \mathcal{B}(t,x)\neq \{ 0\} \textrm{ then } \label{eq:cbf_condition2} \\ &q(\beta^*(t,x),t,x)(u - u^*(t,x)) \leq \alpha(-H(t,x)) - W(t,x) \} \,. \nonumber  
\end{align}
If $\mathcal{B}(t,x) = \{0\}$, then $\boldsymbol\mu_\textrm{rcbf}(t,x) = \mathcal{U}$. We now present the robust extension of Lemma~\ref{prior:predictive_cbf_thm}.

\color{black}
\begin{thm} \label{thm:predictive_cbf}
    Let Assumption~\ref{as:standard} hold. For $H$ as in \eqref{eq:predictive_rcbf}, define $\boldsymbol\mu_\textnormal{\textrm{rcbf}}$ as in \eqref{eq:cbf_condition2} and define $\mathcal{S}_H$ as in \eqref{eq:inner_safe_set}, where $\mathcal{S}_H \equiv \mathcal{S}_H^\textrm{\textnormal{res}}$. Then for any $\alpha \in \mathcal{K}$, the set $\boldsymbol\mu_\textnormal{\textrm{rcbf}}(t,x)$ is nonempty for all $(t,x)\in\mathcal{T}_H$, and any control law $u(t,x)$ that is locally Lipschitz {\color{black}continuous} in $x$ and piecewise continuous in $t$ such that $u(t,x) \in \boldsymbol\mu_\textnormal{\textrm{rcbf}}(t,x), \forall (t,x)\in\mathcal{T}_H$ renders $\mathcal{S}_H$ forward invariant. 
\end{thm}
\color{black}

\color{black}
\begin{pf}\color{black}
    {\color{black}If $\mathcal{B}(t,x) = \{0\}$, then $\boldsymbol\mu_\textrm{rcbf}(t,x) = \mathcal{U}$, which is always nonempty, and $\dot{H}$ in \eqref{eq:etadot_simplified} is nonpositive regardless of the control input $u$. If instead $\mathcal{B}(t,x) \neq \{0\}$,} then the disturbance $w_u$ that maximizes {\color{black}\eqref{eq:etadot_simplified}} has magnitude $w_{u,\textrm{max}}$ and is in the direction of {\color{black}$q(\beta^*(t,x),t,x)$}. Denote this disturbance as $w_{u,\textrm{worst}}\color{black} = \frac{q(\beta^*(t,x),t,x)}{\|q(\beta^*(t,x),t,x)\|} w_{u,\textrm{max}}\color{black}$, and note that this is a known quantity because we assumed at most one nonzero $\beta_c \in \mathcal{B}(t,x)$ (note that if instead we allowed multiple nonzero $\beta_c$, then we could not uniquely define $w_{u,\textrm{worst}}$ here). {\color{black}By Assumption~\ref{as:standard},} the control input $u = u^*(t,x) - w_{u,\textrm{worst}}$ exists in $\mathcal U$ since $u^*$ is restricted to $\mathcal U_w$ in \eqref{eq:noisy_u_set}. {\color{black}This choice of $u$ also renders $\dot{H}(t,x,u,w_u) \leq 0$ in \eqref{eq:etadot_simplified} for any $w_u$, and therefore belongs to $\boldsymbol\mu_\textrm{rcbf}(t,x)$ in \eqref{eq:cbf_condition2} for any $\alpha\in\mathcal{K}$, so $\boldsymbol\mu_\textrm{rcbf}$ is always nonempty regardless of the choice of $\alpha\in\mathcal{K}$.}
    
    \color{black}
    With $W$ as in \eqref{eq:new_W}, any $u(t,x) \in \boldsymbol\mu_\textrm{rcbf}(t,x)$ in \eqref{eq:cbf_condition2} causes $\dot{H}$ in \eqref{eq:etadot_simplified} to satisfy $\dot{H}(t,x,u,w_u) \leq \alpha(-H(t,x))$ regardless of the disturbance $w_u$. Let $\eta(t) = H(t,x(t))$, where $\eta$ is absolutely continuous by Lemma~\ref{lemma:Hpred_derivative_exact} and $\dot{\eta}(t) = \dot{H}(t,x(t),u,w_u) \leq \alpha(-\eta(t))$. Then, given $\eta(t_0) = H(t_0,x(t_0))\leq 0$, Lemma~\ref{prior:nonpositive_lemma} implies that such a control law renders $\eta(t) = H(t,x(t)) \leq 0$ for all $t \in \mathcal{D}$, or equivalently renders $\mathcal{S}_H$ forward invariant. \color{black} \hfill $\blacksquare$
\end{pf}
\color{black}

\color{black}
\begin{cor}\label{cor:is_a_rcbf}
    Suppose the assumptions of Theorem~\ref{thm:predictive_cbf} hold. If additionally $\mathcal{B}(t,x)$ has exactly one element for all $(t,x)\in\mathcal{T}_H$, then $H$ in \eqref{eq:predictive_rcbf} \color{black}is a RCBF on $\mathcal{S}_H$ in \eqref{eq:inner_safe_set} for the system \eqref{eq:model} with $w_x\equiv 0$ {\color{black}for any $\alpha\in\mathcal{K}$}, {\color{black}and \eqref{eq:new_W},\eqref{eq:cbf_condition2} are equivalent to \eqref{eq:max_error},\eqref{eq:valid_rcbf_control}, where}
    \color{black}\begin{gather}
        \partial_t H(t,x) = \frac{\partial h(a_1)}{\partial \lambda_t} + \frac{\partial h(a_1)}{\partial \lambda_x} \theta(\beta_c,t,x) \,, \label{eq:partial1} \\ 
        \nabla H(t,x) = \frac{\partial h(a_1)}{\partial \lambda_x} \Theta(\beta_c,t,x) \,. \label{eq:partial2}
    \end{gather}
    and where $a_1$ is as in \eqref{eq:predictive_rcbf} and $\beta_c = \mathcal{B}(t,x)$.
\end{cor}

\begin{pf}
    \color{black}
    Under this stricter condition on $\mathcal{B}(t,x)$, the function $H$ in \eqref{eq:predictive_rcbf} is now continuously differentiable \cite[Lemma 1]{Predictive_CBF}. By the same argument as in Theorem~\ref{thm:predictive_cbf}, the control input $u = u^*(t,x) - w_{u,\textrm{worst}}$ must exist in $\mathcal{U}$, and makes $\dot{H}$ \color{black}at most 0 independent of the actual disturbance $w_u$. Thus, condition \eqref{eq:rcbf_definition} is satisfied for any $\alpha\in\mathcal{K}$, so $H$ is a RCBF on $\mathcal{S}_H$ {\color{black}for any $\alpha\in\mathcal{K}$}. 
    \regularversion{{\color{black}The proof of \eqref{eq:partial1}-\eqref{eq:partial2} follows from \eqref{step5}; for a complete proof, see \cite[Cor.~\ref{cor:is_a_rcbf}]{extended}.}\hfill $\blacksquare$}
    
\extendedversion{
    Next, since $H$ is continuously differentiable, we can now properly define its partial derivatives. We proved in Theorem~\ref{thm:predictive_cbf} that $\frac{d h(a_1)}{d \beta_c}\frac{d \beta_c}{dt} = 0$, where now $\beta_c(t,x) \equiv \mathcal{B}(t,x)$. Expanding this, it follows that
    \begin{align}
        0 =& \left( \frac{\partial h(a_1)}{\partial \lambda_t} + \frac{\partial h(a_1)}{\partial \lambda_x}\frac{\partial \chi(\beta_c,t,x)}{\partial \lambda_\beta} \right) \nonumber \\ &\;\;\;\; \cdot \(\frac{\partial \beta_c(t,x)}{\partial t} + \frac{\partial \beta_c(t,x)}{\partial x} \frac{dx}{dt}\right) \label{extra_step1}
    \end{align}
    Note that \eqref{extra_step1} must hold for any $\frac{dx}{dt}$, so we conclude that $ \left( \frac{\partial h(a_1)}{\partial \lambda_t} + \frac{\partial h(a_1)}{\partial \lambda_x}\frac{\partial \chi(\beta_c,t,x)}{\partial \lambda_\beta} \right) \frac{\partial \beta_c(t,x)}{\partial x} \equiv 0$. It follows that $\left( \frac{\partial h(a_1)}{\partial \lambda_t} + \frac{\partial h(a_1)}{\partial \lambda_x}\frac{\partial \chi(\beta_c,t,x)}{\partial \lambda_\beta} \right)\frac{\partial \beta_c(t,x)}{\partial t} \equiv 0$ too. That is, we have extended the result on the sensitivity of $h$ to the total derivative of $\beta_c$ from Theorem~\ref{thm:predictive_cbf} to the sensitivity of $h$ to the partial derivatives of $\beta_c$ as well. This allows us to separate \eqref{step5} into the desired partial derivatives.
    
    Next, the partial derivatives of \eqref{eq:predictive_rcbf} are equivalent to the partial derivatives of \eqref{eq:Hequiv}, which are
    \begin{align}
        \partial_t H(t,x) =& \frac{\partial h(a_1)}{\partial \lambda_t}\(1+\frac{\partial \beta_c(t,x)}{\partial t}\) + \frac{\partial h(a_1)}{\partial \lambda_x} \( \frac{\partial \chi(\beta_c, t, x)}{\partial \lambda_{t_0}}  \nonumber \right. \\ &\;\; \left. + \frac{\partial \chi(\beta_c, t, x)}{\partial \lambda_\beta}\frac{\partial \beta_c(t,x)}{\partial t} \right) \label{extra_step2} \,, \\
        \nabla H(t,x) =& \frac{\partial h(a_1)}{\partial \lambda_t} \frac{\partial \beta_c(t,x)}{\partial x} + \frac{\partial h(a_1)}{\partial \lambda_x} \( \frac{\partial \chi(\beta_c, t, x)}{\partial \lambda_{x_0}} \nonumber \right. \\  &\;\; \left. + \frac{\partial \chi(\beta_c, t, x)}{\partial \lambda_\beta}\frac{\partial \beta_c(t,x)}{\partial x} \) \label{extra_step3} \,.
    \end{align}
    Cancelling the zero terms identified above, \eqref{extra_step2}-\eqref{extra_step3} simplify to
    \begin{align}
        \partial_t H(t,x) =& \frac{\partial h(a_1)}{\partial \lambda_t} + \frac{\partial h(a_1)}{\partial \lambda_x} \frac{\partial \chi(\beta_c, t, x)}{\partial \lambda_{t_0}} \,, \\
        \nabla H(t,x) =& \frac{\partial h(a_1)}{\partial \lambda_x} \frac{\partial \chi(\beta_c, t, x)}{\partial \lambda_{x_0}} \,,
    \end{align}
    which are equivalent to \eqref{eq:partial1}-\eqref{eq:partial2}. \hfill {\color{black}$\blacksquare$}
}
\end{pf}
\color{black}

\extendedversion{
% {\color{black}
\renewcommand{\thethm}{on Corollary}
\begin{rem}
    The extra condition that $\mathcal{B}(t,x)$ have exactly one element in Corollary~\ref{cor:is_a_rcbf} is a technicality, but is required to ensure that $H$ in \eqref{eq:predictive_rcbf} is continuously differentiable, which was a condition of Definition~\ref{def:rcbf}. 
    If we had instead defined RCBFs using directional derivatives, then we would not need to make this distinction.
\end{rem}
% }
\addtocounter{thm}{-1}
\renewcommand{\thethm}{\arabic{thm}}
}

\begin{rem} \label{rem:predictive_unmatched}
    The authors are not aware of a method by which Theorem~\ref{thm:predictive_cbf} {\color{black}and Corollary~\ref{cor:is_a_rcbf}} can be extended to the case of {\color{black}$w_{x}\not\equiv 0$} with guaranteed safety. In this case, we refer readers to the robustness method in {\color{black}\cite[Prop. 5]{Robust_CBFs}}. 
    This could require expansion of the control set $\mathcal{U}$.
\end{rem}

In other words, if the undisturbed trajectory described by $\chi(\beta,t,x(t))$ remains safe for all future time, then there always exists a safe trajectory from $(t,x(t))$ regardless of the actual disturbance (for matched disturbances only). It is trivial to show {\color{black}that} $H$ in \eqref{eq:predictive_rcbf} satisfies $H(t,x)\geq h(t,x),\forall (t,x)\in\mathcal{D}\times\reals^n$, so the inner safe set $\mathcal{S}_H$ in Theorem~\ref{thm:predictive_cbf} is a subset of the safe set $\mathcal{S}$. 
{\color{black}In the prior work \cite{CDC21}, Lemma~\ref{lemma:constant_a} was presented as a special case of Lemma~\ref{prior:predictive_cbf_thm}. In this paper, the assumptions of Theorems~\ref{thm:constant_noise_cbf},\ref{thm:variable_cbf} are different from those of Theorem~\ref{thm:predictive_cbf}, but Theorem~\ref{thm:predictive_cbf} is still the most generally applicable {\color{black}method of finding a set $\mathcal{S}_H$} because there may exist $u^*$ satisfying 
{\color{black}Assumption~\ref{as:standard}}
even when there is no $a_\textrm{max}$ or $\Phi$ satisfying the assumptions of Theorems~\ref{thm:constant_noise_cbf} or \ref{thm:variable_cbf}, respectively.}

Note that satisfying the assumptions on $\mathcal{B}(t,x)$ {\color{black}in Assumption~\ref{as:standard}} could be easy or challenging depending on the system and choice of $u^*$, as will be elaborated upon in application in Section~\ref{sec:spacecraft_functions}.
{\color{black}The need to verify {\color{black}these assumptions} is perhaps the primary limitation of Theorem~\ref{thm:predictive_cbf}, and is a topic of future study. In the case of relative degree $r=2$ or $r=3$, a sufficient condition for this assumption to hold is the existence of $\gamma \in\reals_{<0}$ such that $h^{(r)}(t,x,u^*(t,x),w_u) \leq \gamma, \forall (t,x)\in\mathcal{T}, \forall w_u \in \mathcal{W}$.}

Applying Theorem~\ref{thm:predictive_cbf} also requires that we know a control law $u^*$ that generally drives the trajectories $\chi$ towards safe states. For a system with relative-degree $r$, one obvious choice of control law is
\begin{equation}
    u^*_\textrm{opt}(t,x) = \argmin_{u\in \mathcal{U}_w} \nabla h^{(r-1)}(t,x) g(t,x) u \,. \label{eq:sample_control_law}
\end{equation}
Certain known phenomena of system trajectories might also motivate other control laws, such as the control law $u^*_\textrm{orth}$ in Section~\ref{sec:spacecraft_functions}, inspired by satellite orbits.

Of the methods presented so far, Theorem~\ref{thm:predictive_cbf} is the most computationally intensive. Unlike the previous methods, the ODE in \eqref{eq:ode_with_noise} rarely simplifies to explicit algebraic expressions. That said, the simulations in Section~\ref{sec:spacecraft_functions} show that the computation costs are reasonable on a 6-dimensional system. 
In practice, we also need an upper bound on the amount of time to propagate \eqref{eq:ode_with_noise} by which time all maximizers of \eqref{eq:predictive_rcbf} will be guaranteed to have occurred, though computation of such a bound is system-specific and not addressed in this paper.

\section{Hysteresis-Switched Control Barrier Functions} \label{sec:switching}

% \subsection{Motivation}

This section develops a condition on the control input that establishes set {\color{black}forward} invariance similar to Lemma~\ref{prior:rcbf_invariance}, but which places fewer constraints on the system trajectories. The core idea of this section is that, as we will show, the RCBF condition $u\in \boldsymbol\mu_\textrm{rcbf}(t,x)$ in \eqref{eq:valid_rcbf_control} does not need to be enforced everywhere in ${\color{black}\mathcal{S}_H^\textrm{res}}$ to ensure forward invariance of ${\color{black}\mathcal{S}_H^\textrm{res}}$, and we will develop a systematic way to relax the conditions in Lemma~\ref{prior:rcbf_invariance}. 

{\color{black}Recall} that the theorems in Section~\ref{sec:new_cbfs} {\color{black}showed that the presented functions $H$ are RCBFs} for any $\alpha\in\mathcal{K}$. Thus, $\alpha$ is a free parameter, though $\alpha$ still impacts system performance by its role in defining $\boldsymbol\mu_\textrm{rcbf}$ in \eqref{eq:valid_rcbf_control}{\color{black},\eqref{eq:cbf_condition2}}.
Near the boundary of the set ${\color{black}\mathcal{S}_H^\textrm{res}}$, the function $\alpha$ works to ensure invariance of ${\color{black}\mathcal{S}_H^\textrm{res}}$ by bounding the rate at which the system approaches this boundary. However, in the interior of the ${\color{black}\mathcal{S}_H^\textrm{res}}$, the bound provided by $\alpha$ is arbitrary (since $\alpha$ is a free parameter) and can prevent the system from following otherwise safe trajectories. For example, consider the CBF in \eqref{eq:constant_abs_cbf}. The CBF condition for this function becomes
\begin{equation}
    \ddot{h}(x,u) \leq -\frac{\dot{h}(x)}{|\dot{h}(x)|} a_{\max} + \frac{\alpha(-H(x))}{|\dot{h}(x)|}a_{\max} \label{eq:example_high_relative_degree}
\end{equation}
Suppose $h,\dot{h},\ddot{h}$ represent position, velocity, and acceleration, respectively. {\color{black}If} an agent is far away from the boundary of ${\color{black}\mathcal{S}_H^\textrm{res}}$ (i.e. $H(x) \ll 0$ {\color{black}and $h(x) \ll 0$}) and moving quickly (i.e. $|\dot{h}| \gg 0$), then the acceleration $\ddot{h}$ may be constrained even though the state is far from the boundary of ${\color{black}\mathcal{S}_H^\textrm{res}}$. This is a consequence of how all CBFs {\color{black}for high-relative-degree constraint functions} are necessarily functions of the constraint function derivatives.
Thus, we seek to remove this unnecessary constraint when $H$ is far less than zero.

\begin{figure}
    \centering
    \includegraphics[width=\columnwidth]{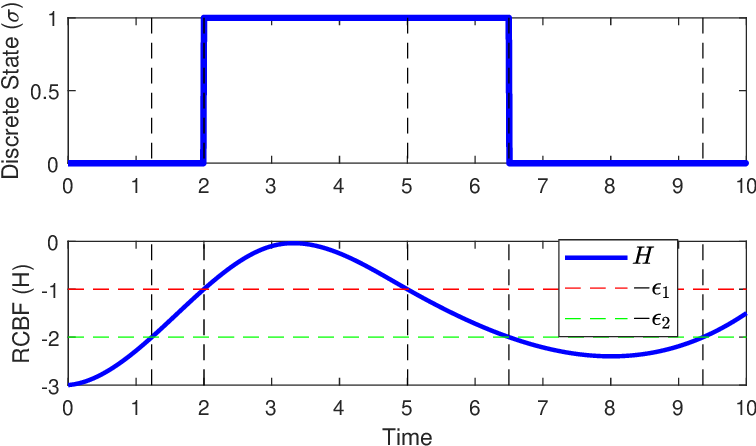}
    \caption{A visualization of state-dependent hysteresis-switching in \eqref{eq:switching_logic}.}
    \label{fig:demo_switch}
\end{figure}

To this end, we introduce a discrete state $\sigma\in\{0,1\}$, which captures whether the state is near the boundary of $\mathcal{S}_H$. We say a particular RCBF condition is \emph{active} if $\sigma=1$, and \emph{inactive} if $\sigma=0$. If there are multiple RCBFs, then one would construct $\sigma_i, i=1,2,\cdots$ for each RCBF. Define $\epsilon_2 > \epsilon_1 \geq 0$ and at a time instant $t$, let
\begin{equation}
    \sigma(t) = \begin{cases} 0 & H(t, x(t)) \leq -\epsilon_2 \\
        1 & H(t, x(t)) \geq -\epsilon_1 \\
        \sigma(t^-) & \textrm{otherwise}
    \end{cases} \,. \label{eq:switching_logic}
\end{equation}
where $\sigma(t_0) = 0$, and $t^-$ denotes the time instant immediately preceding $t$. That is, $\sigma$ exhibits hysteresis, as visualized in the example in Fig.~\ref{fig:demo_switch}; the constraint becomes active ($\sigma \rightarrow 1$) when $H$ first exceeds some tolerance ($H\geq -\epsilon_1$), and remains active until the state is far away from the boundary of $\mathcal{S}_H$ ($H \leq -\epsilon_2)$. We can then use this discrete state to choose between two control laws depending on whether the constraint is active or inactive. The result is a switched control law with state-dependent switching, where we say a \textit{switch} occurs when $\sigma(t)$ changes between discrete states. We denote a switch to active as $\sigma\rightarrow 1$ and a switch to inactive as $\sigma\rightarrow 0$. Since trajectories of $x$ are continuous and $H$ is continuous, the state can never leave {\color{black}$\mathcal{S}_H^\textrm{res}\subseteq\mathcal{S}$} without the corresponding RCBF $H$ becoming active. Thus, we establish safety under such a switched control law in the following theorem.

\begin{thm} \label{thm:switched_control}
    If $H:\mathcal{D}\times\reals^n\rightarrow\reals$ is a RCBF on $\mathcal S_H$ in \eqref{eq:inner_safe_set}, then any control law $u(t, x, \sigma)$ with $\sigma$ as in \eqref{eq:switching_logic} that is locally Lipschitz {\color{black}continuous} in $x$ and piecewise continuous in $t$ such that $u(t,x,1) \in \boldsymbol\mu_{\textrm{\textnormal{rcbf}}}(t,x), \forall (t,x) \in \{(t,x) \in \mathcal{T}_H \mid H(t,x) \geq -\epsilon_2 \}$ will render $\mathcal{S}_H$ forward invariant.
\end{thm}
\begin{pf}
    Under the above assumptions on $u$, solutions $x(t), t \in \mathcal D$ to dynamics \eqref{eq:model} are unique {\color{black}\cite[Thm. 54]{Sontag}} and absolutely continuous {\color{black}\cite[Eq. C.2]{Sontag}} for all $t \in \mathcal D$. Therefore, $\eta(t) = H(t,x(t))$ is absolutely continuous, and regardless of the disturbances $w_u,w_x$, $\dot{\eta}$ satisfies $\dot{\eta}(t) = \dot{H}(t,x(t),u,w_u,w_x) \leq \alpha(-\eta(t))$ for all $t \in \mathcal{D}_1 \triangleq \{t\in\mathcal{D} \mid \sigma(t) = 1\}$. Thus, by Lemma~\ref{prior:nonpositive_lemma}, if $H(t_0,x(t_0))\leq 0$, then $H(t,x(t)) = \eta(t) \leq 0, \forall t \in \mathcal{D}_1$. Also, by construction, $H(t,x(t))\leq -\epsilon_1 \leq 0, \forall t \in \mathcal{D}\setminus \mathcal{D}_1$. Thus, $H(t,x(t))\leq 0, \forall t \in \mathcal{D}$, which implies $\mathcal{S}_H$ is forward invariant. \hfill $\blacksquare$
\end{pf}

Note that one can come to a similar conclusion as in Theorem~\ref{thm:switched_control} from \cite[Thm. 3]{Glotfelter2}, but by using a hysteresis switching logic, we avoid the need for differential inclusions and non-smooth analysis. Such switching prevents sliding modes with chattering control inputs that may be introduced by differential inclusions, and thus results in more realistic actuator behavior, while still providing the benefits of allowing potentially discontinuous controllers. 
{\color{black}A related approach for smooth switching based on the value of $h$ rather than $H$ is presented in \cite{robust_hocbfs}, but this approach requires $\mathcal{U}=\reals^m$. Compared to \cite{First_CBF,multi_unicycle_2021}, the switching approach in Theorem~15 still utilizes the RCBF condition in \eqref{eq:valid_rcbf_control}, and thus can be included in optimization-based controllers, as is done in Section~\ref{sec:spacecraft_functions}.}
Note that if $\epsilon_1 = \infty$, then Theorem~\ref{thm:switched_control} reduces to Lemma~\ref{prior:rcbf_invariance}.

\begin{rem}
    Note that the argument used to prove invariance of $\mathcal{S}_H^\textrm{\textnormal{res}}$ in Lemma~\ref{lemma:absolute_value} is unaffected by the addition of switching, since it holds that $\dot{h}(x) \leq 0$ on $\mathcal{S}_H^\textrm{\textnormal{com}}$ regardless of the control input. Thus, $\mathcal{S}_H$ may be replaced by $\mathcal{S}_H^\textrm{\textnormal{res}}$ in Theorem~\ref{thm:switched_control} {\color{black}for the RCBFs \eqref{eq:constant_noise_cbf} and \eqref{eq:variable_cbf}}. {\color{black}Similarly, $H$ in Theorem~\ref{thm:predictive_cbf} may be used even if the stricter conditions of Corollary~\ref{cor:is_a_rcbf} do not hold.}
\end{rem}

For practical implementation, it is often desirable to tune a controller to reduce the number of switches of the discrete state, which could cause jumps in the control input that wear out the actuators. This tuning can be done directly in the RCBF development by proper choice of the class-$\mathcal{K}$ function $\alpha$ and the switching tolerances $\epsilon_1,\epsilon_2$. To this end, we propose the function 
\begin{equation}
    \alpha_r(\lambda, t, x) \triangleq \frac{W(t,x)\lambda}{\epsilon_1} \label{eq:chosen_alpha}
\end{equation}
in place of $\alpha$ in \eqref{eq:rcbf_definition},\eqref{eq:valid_rcbf_control},{\color{black}\eqref{eq:cbf_condition2}} and the choices $\epsilon_2 > 2 \epsilon_1$ and $\epsilon_1 > 0$. The reasoning for {\color{black}choosing $\epsilon_2 > 2\epsilon_1$} is that the disturbances $w_u, w_x$ could be helpful or harmful to safety. If $\sigma=1$ and the disturbance is temporarily helpful to safety {\color{black}(i.e. causing $H$ to decrease)}, we wish to avoid a switch $\sigma \rightarrow 0$ at states where a disturbance harmful to safety {\color{black}(i.e. causing $H$ to increase)} could quickly cause another switch back to $\sigma\rightarrow 1$. 
{\color{black}To this end, suppose that the control input satisfies the RCBF condition in \eqref{eq:valid_rcbf_control} or \eqref{eq:cbf_condition2} with equality. Then the set $\mathcal{S}_f(t) \triangleq \{x\in \mathcal{S}_H^\textrm{res}(t) \mid \dot{H}(t,x)=0\}$ is asymptotically stable. Next, suppose $\alpha = \alpha_r$ as in \eqref{eq:chosen_alpha}. Then $\dot{H}$ is given by}
\begin{align}
    %\color{black}\dot{H}(\cdot,u,w_u,w_x) = \dot{H}(\cdot,u,0,0) + W(\cdot) + \nabla H(\cdot) (g(\cdot) w_u + w_x) - W(\cdot) \color{black} \nonumber \\
    %
    \color{black} \dot{H}(&\color{black}\cdot) = \partial_t H(\cdot) + \nabla H(\cdot)\big( f(\cdot) + g(\cdot) (u + w_u) + w_x\big) \nonumber \\
    %
    %\dot{H}(\cdot) 
    &\color{black}\overset{\eqref{eq:valid_rcbf_control}}{=} \alpha(-H(\cdot)) - W(\cdot) + \nabla H(\cdot) \big(g(\cdot)w_u + w_x\big) \nonumber \\
    &\color{black}\overset{\eqref{eq:chosen_alpha}}{=} \color{black} -\frac{W(\cdot) H(\cdot)}{\epsilon_1} + \underbrace{\nabla H(\cdot) (g(\cdot) w_u + w_x)}_{=W_\textrm{real}(t,x,w_u,w_x)} - W(\cdot) \label{eq:alpha_r_implemented}
\end{align}
{\color{black}where $W_\textrm{real}$ captures the effect on $\dot{H}$ of the disturbances that occur online.}
If the {\color{black}online} disturbances $w_u, w_x$ are maximally harmful to safety (i.e. tending to increase $\dot{H}$), then $W_\textrm{real} = W$ in \eqref{eq:alpha_r_implemented} and {\color{black}$\mathcal{S}_f(t) \equiv \{x\in \mathcal{S}_H^\textrm{res}(t) \mid H(t,x)=0\}$, so} {\color{black}$H$ converges to 0 asymptotically.} 
If instead $w_u,w_x$ are maximally helpful to safety ({\color{black}i.e.} tending to decrease $\dot{H}$), 
then $W_\textrm{real} = -W$ and {\color{black}$\mathcal{S}_f(t) \equiv \{x\in \mathcal{S}_H^\textrm{res}(t) \mid H(t,x)=-2\epsilon_1\}$, so} {\color{black}$H$ converges to $-2\epsilon_1$ asymptotically.} 
Thus, choosing $\epsilon_2 > 2\epsilon_1$ {\color{black}in the switching conditions} prevents a switch induced purely by the disturbance. Instead, switches will occur when the system objectives drive the trajectory closer to or further from the boundary of {\color{black}$\mathcal S_H^\textrm{res}$}{\color{black}, i.e. when the control input changes between satisfying \eqref{eq:valid_rcbf_control},\eqref{eq:cbf_condition2} with and without equality. Finally,} if the disturbance is zero, then {\color{black}$\mathcal{S}_f(t) \equiv \{x\in \mathcal{S}_H^\textrm{res}(t) \mid H(t,x)=-\epsilon_1\}$, so} {\color{black}$H$ will converge to $-\epsilon_1$.} 
{\color{black}That is, {\color{black}using $\alpha=\alpha_r$ in \eqref{eq:chosen_alpha} allows us to 
tune how closely the closed-loop trajectories approach the boundary of $\mathcal{S}_H^\textrm{res}$ as a function of the disturbances}.} 
Note that $\alpha_r$ does not belong to class-$\mathcal{K}$ {\color{black}(as required by Lemma~\ref{prior:rcbf_invariance} and Theorem~\ref{thm:switched_control})}, so the following corollary shows that {\color{black}$\alpha_r$ can} still be used for safety.

\begin{cor} \label{cor:chosen_alpha}
    Suppose $g(t,x)$ and $\nabla H(t,x)$ are bounded for all $(t,x)\in \mathcal{T}_H$. Then Lemma~\ref{prior:rcbf_invariance} and Theorem~\ref{thm:switched_control} hold when $\boldsymbol\mu_\textrm{\textnormal{rcbf}}(t,x)$ in \eqref{eq:valid_rcbf_control} is defined with $\alpha = \alpha_r$ as in \eqref{eq:chosen_alpha}. {\color{black}Similarly, if $q(\beta^*(t,x),t,x)$ is bounded for all $(t,x)\in \mathcal{T}_H$, then  Theorem~\ref{thm:predictive_cbf} holds when  \eqref{eq:cbf_condition2} is defined with $\alpha = \alpha_r$.}
\end{cor}
\begin{pf}
    Under the above assumptions, $W(t,x)$ {\color{black}in \eqref{eq:max_error} and \eqref{eq:new_W}} is upper bounded. Let $w_m$ denote this bound. Then $\alpha_r(\lambda, t, x) \leq \bar{\alpha}_r(\lambda) \triangleq w_m \lambda / \epsilon_1$, where $\bar{\alpha}_r \in \mathcal{K}$. Thus, a control input $u\in\boldsymbol\mu_\textrm{rcbf}(t,x)$ yields $\dot{\eta}(t) \leq \bar{\alpha}_r(-\eta(t))$ as in Theorem~\ref{thm:switched_control}, so the result of Theorem~\ref{thm:switched_control} still holds.
    By the same argument, Lemma~\ref{prior:rcbf_invariance} {\color{black}and Theorem~\ref{thm:predictive_cbf}} hold as well.
    \hfill $\blacksquare$
\end{pf}

Thus, we have proposed a method of regulating safety that allows for switched controllers that mitigate the potentially undesirable constraints following from high-relative-degree constraint functions. 
{\color{black}Using the position/velocity/acceleration interpretation of $h$, $\dot{h}$, $\ddot{h}$, respectively, this switching approach allows an agent to 
{\color{black}choose control inputs without considering the CBF condition when} 
in the interior of its safe set and then to decelerate approximately along the surface $\mathcal{S}_f(t)$ as it approaches an unsafe region. This} approach also allows for a provably safe means of adding and removing RCBFs over time, for instance, as an agent explores an unknown environment and identifies obstacles.

\section{Spacecraft Simulation and Discussion} \label{sec:spacecraft_functions}

\subsection{Spacecraft Dynamics}

In this section, we apply the previously presented RCBFs and hysteresis-switching strategy to satellite dynamics and demonstrate these approaches in simulation. The satellite state is $x = \[ r\transpose\; v\transpose \]\transpose$, with dynamics
\begin{equation}
    \dot{x} = \begin{bmatrix} \dot{r} \\ \dot{v} \end{bmatrix} = \underbrace{\begin{bmatrix} v \\ f_\mu(t,r) \end{bmatrix}}_{f(t,x)} + \underbrace{\begin{bmatrix} 0_{3\times 3} \\ I_{3\times 3} \end{bmatrix}}_{g(t,x)} (u + w_u) + \begin{bmatrix} w_x \\ 0 \end{bmatrix} \label{eq:spacecraft_dynamics}
\end{equation}
for {\color{black}position and velocity} $r, v \in\reals^3$ 
and gravitational force $f_\mu:\mathcal{D}\times\reals^3\rightarrow\reals^3$. For compactness, we assume $w_x$ only acts on the $\dot{r}$ equation (any effects on $\dot{v}$ can be grouped with $w_u$). In this system, the matched disturbance $w_u$ could represent unmodelled forces like higher-order gravity effects or solar radiation pressure, while the unmatched disturbance could represent filtered sensor updates. Suppose the satellite mass $m$ is approximately constant and the satellite contains 6 orthogonal thrusters (or fewer thrusters capable of changing orientation sufficiently fast) capable of outputting a continuously variable thrust in $[0, u_\textrm{max}m]$ for some $u_\textrm{max}>0$. Then the control set is $\mathcal U = \{u\in\reals^3 \mid ||u||_\infty \leq u_\textrm{max} \}$ and represents the satellite allowable acceleration.
The following examples compute trajectories entirely online (i.e. no advance path-planning) {\color{black} and} assume no global velocity bound. 

\subsection{Setup for Safety}\label{sec:safety_setup}

{\color{black}The \extendedversion{first }proposed mission centers around a flyby of the asteroid Ceres\footnote{\color{black}All simulation code may be found at \url{https://github.com/jbreeden-um/phd-code/}}. A second simulation around the asteroid Eros \regularversion{can be found in the extended version \cite{extended}}\extendedversion{is also included in Section~\ref{sec:results_prox}}. The safe set $\mathcal{S}$ is the set of states with positions $r$ sufficiently far from the asteroid and arbitrary velocities $v$, while the sets $\mathcal{S}_H$, $\mathcal{S}_H^\textrm{res}$ add restrictions on $v$. Safety is encoded by the constraint function}
\begin{equation}
    h_c(t,x) \triangleq \rho - ||r - r_c(t)|| \,,
\end{equation}
where $r_c\in\reals^3$ is the point to be avoided and $\rho$ is the minimum allowable distance. Suppose this point has known velocity $v_c$ and acceleration $u_c$. 
The derivatives of $h_c$ following from \eqref{eq:def_hw} are as follows.
\begin{align}
    \dot{h}_{w,c}(t,x) = -&\frac{(r-r_c(t))\transpose(v-v_c(t))}{\|r - r_c(t)\|} + w_{x,\textrm{max}} \\
    \ddot{h}_{w,c}(t,\hspace{-0.5pt}x,\hspace{-0.5pt}u,\hspace{-0.5pt}w_u) \hspace{-1pt}&=\hspace{-1pt} -\frac{(r\hspace{-1pt}-\hspace{-1pt}r_c(t))\transpose(f_\mu(t,r) \hspace{-1pt}+\hspace{-1pt} u \hspace{-1pt}+\hspace{-1pt} w_u \hspace{-1pt}-\hspace{-1pt} u_c(t))}{\|r-r_c(t)\|} \nonumber \\ & \;\;\;\;\;\;\;\; - \frac{\| (r-r_c(t))^\times (v-v_c(t)) \|^2}{\|r - r_c(t)\|^3} \label{eq:hwc_ddot}
\end{align}
For this mission, we assume {\color{black}that} only low-thrust actuators {\color{black}are available}, meaning that $u_\textrm{max} \ll \|f_\mu\|$ in the vicinity of the asteroid. Specifically, let $u_\textrm{max} = (10)^{-4}$ m/s$^2$, which is approximately the peak acceleration achievable by the DAWN spacecraft halfway through its mission\footnote{Ref: \url{https://solarsystem.nasa.gov/missions/dawn/technology}}, or a modern SmallSat ion thruster\footnote{E.g. Busek Bit-3 on a 12 kg CubeSat: \url{http://www.busek.com/technologies__ion.htm}}. The radius of Ceres is approximately $\rho_\textrm{Ceres}=476000$ m, and 
gravitational acceleration near Ceres is given by
\begin{equation}
    f_{\mu,c}(t,r) = -\frac{\mu (r-r_c)}{\|r-r_c(t)\|^3} \,, \label{eq:central_gravity}
\end{equation}
where $\mu = 6.26325(10)^{10}$ is fixed.

\begin{table}[]
    \color{black}
    \centering
    \begin{tabular}{c|c|c|c|c}
        RCBF & Theorem & Form & $\rho$ & Parameter \\ \hline
        $H_1^A$ & \ref{thm:constant_noise_cbf} & \eqref{eq:constant_noise_cbf} & $3.63(10)^7$ & 
        $a_\textrm{max}$ in \eqref{eq:implemented_a_max} \\
        $H_2^A$ & \ref{thm:variable_cbf} & \eqref{eq:variable_cbf} & $3.21(10)^7$ & $\Phi_c$ in \eqref{eq:implemented_phi} \\
        $H_3^A$ & \ref{cor:is_a_rcbf} & \eqref{eq:predictive_rcbf} & $2.50(10)^7$ & $u^*_\textrm{rad}$ in \eqref{eq:u_ball}\\
        $H_4^A$ & \ref{thm:predictive_cbf} & \eqref{eq:predictive_rcbf} & $4.76(10)^5$ & $u^*_\textrm{orth}$ in \eqref{eq:u_orth}
    \end{tabular}
    \caption{A summary of the 4 RCBFs tested and the parameters used\extendedversion{{} for simulations around Ceres}}
    \label{tab:cbfs_summary}
\end{table}

{\color{black}Simulations were run with 4 different RCBFs, summarized in Table~\ref{tab:cbfs_summary} and detailed as follows.}
{\color{black}First, we apply Theorem~\ref{thm:constant_noise_cbf}. Substituting the dynamics \eqref{eq:spacecraft_dynamics} into \eqref{eq:a_max0}, the parameter $a_\textrm{max}$ is given by}
\begin{equation}
    a_\textrm{max} = u_\textrm{max} - w_{u,\textrm{max}} - \sup_{t\in\mathcal{D}, x\in\mathcal{S}(t)} \| f_\mu(t,r) \|. \label{eq:implemented_a_max}
\end{equation}
One possible RCBF{\color{black}, which we denote $H_1^A$,} is then given by \eqref{eq:constant_noise_cbf} with $h_c, \dot{h}_{w,c}$ in place of $h,\dot{h}_w$, respectively. 
Because $u_\textrm{max} \ll \|f_\mu\|$ near the asteroid, we must choose $\rho$ large enough that $a_\textrm{max}$ in \eqref{eq:implemented_a_max} is positive. Specifically, we chose $\rho$ as $\rho^A_1 = 3.63(10)^7 \textrm{ m}$, which along with $w_{u,\textrm{max}}=5(10)^{-6} \textrm{ m/s}^2$ leads to $a_\textrm{max} = 4.55(10)^{-5} \textrm{ m/s}^2$ in \eqref{eq:implemented_a_max}.

{\color{black}Second, we} note that $\ddot{h}_{w,c}$ {\color{black}in \eqref{eq:hwc_ddot}} satisfies
\begin{multline}
    \max_{\|w_u\|\leq w_{u,\textrm{max}}}\inf_{u\in \mathcal U} \ddot{h}_{w,c}(t,x,u,w_u) \\ \leq \frac{\mu}{\| r-r_c(t) \|^2} - u_\textrm{max} + w_{u,\textrm{max}} . \label{eq:inf_hddot}
\end{multline}
{\color{black}Since} $\| r-r_c(t)\| = \rho - h_c(t,x)$, %so
the right hand side of \eqref{eq:inf_hddot} can be written as a function $\phi{\color{black}_c}(h_c(t,x))$ only dependent on $h_c$ and constants. 
{\color{black}The anti-derivative of $\phi_c$ is then}
\begin{equation}
    \Phi_{\color{black}c}(\lambda) = \frac{\mu}{\rho - \lambda} + (w_{u,\textrm{max}} - u_\textrm{max}) \lambda \label{eq:implemented_phi} \,.
\end{equation}
{\color{black}Applying Theorem~\ref{thm:variable_cbf},} a second possible RCBF{\color{black},which we denote $H_2^A$,} is then given by \eqref{eq:variable_cbf} with $h_c, \dot{h}_{w,c}{\color{black},\Phi_c}$ in place of $h,\dot{h}_w{\color{black},\Phi}$, respectively. {\color{black}Note that Theorem~\ref{thm:variable_cbf} requires that} the right hand side of \eqref{eq:inf_hddot} is always {\color{black}negative, so} we chose $\rho$ as $\rho^A_2 = 3.21(10)^7 \textrm{ m}$. {\color{black}In this system,} the function $\Phi{\color{black}_c}$ {\color{black}in \eqref{eq:implemented_phi}} follows directly from $f_\mu$ being a potential force, and while that is not necessary in general, it may be hard to find a function {\color{black}$\Phi$} satisfying the conditions of Theorem~\ref{thm:variable_cbf} for systems without known expressions for potential energy. On the other hand, it is comparatively easy to find a constant $a_\textrm{max}$ (or show than none exists) {\color{black}using \eqref{eq:a_max0}} even for complex systems.

\extendedversion{Note that we also could have used the equivalent constraint function $h_{alt}(t,x) \triangleq \rho^2 - \| r - r_c(t)\|^2$. However, no valid $a_\textrm{max}$ or $\Phi$ exists for $h_{alt}$, so one may need to choose the constraint function carefully as well. Intuitively, $h_c$ represents position inside a potential field, while $h_{alt}$ has no physical interpretation in the context of the dynamics in \eqref{eq:spacecraft_dynamics}.}

{\color{black}Third,} we apply Theorem~\ref{thm:predictive_cbf} {\color{black}for two different control laws $u^*$}. Note that the constructions of both $a_\textrm{max}$ in \eqref{eq:implemented_a_max} and $\Phi$ in \eqref{eq:implemented_phi} ignore the effect of the second term of $\ddot{h}_{w,c}$ in \eqref{eq:hwc_ddot}, which is always nonpositive (i.e. helpful to safety), but not amenable to simple RCBF formulas. Theorem~\ref{thm:predictive_cbf} allows us to account for this term as well, and thus decrease conservatism. 
First, define the control law
\begin{equation}
    u_\textrm{rad}^*(t,x) \triangleq \argmin_{u\in\mathcal{U}_w} -(r - r_c(t)) u \,, \label{eq:u_ball}
\end{equation}
where $\mathcal{U}_w$ is given in \eqref{eq:noisy_u_set}. This is a special case of $u^*_\textrm{opt}$ in \eqref{eq:sample_control_law}. In addition to capturing the effect of the second term of \eqref{eq:hwc_ddot}, using Theorem~\ref{thm:predictive_cbf} with this control law also captures how the spacecraft has more control authority when $r-r_c$ has components in the direction of more than one thruster. %The
{\color{black}A third possible} RCBF{\color{black}, which we denote $H_3^A$}, is then given by \eqref{eq:predictive_rcbf} with $h_c, u^*_\textrm{rad}$ in place of $h, u^*$, respectively.

Lastly, define the control law
\begin{equation}
    u_\textrm{orth}^*(t,x) \triangleq \argmin_{u\in\mathcal{U}_w} v_\textrm{orth}(t,x)u \,, \label{eq:u_orth}
\end{equation}
where
\begin{multline}
    v_\textrm{orth}(t,x) \triangleq v-v_c(t)  - \frac{(r-r_c(t))\transpose(v-v_c(t))}{\|r-r_c(t)\|^2}(r-r_c(t)) \,. \nonumber
\end{multline}
While $u^*_\textrm{rad}$ thrusts away from the asteroid, the control law $u^*_\textrm{orth}$ thrusts tangential to the asteroid, thereby making the second term of \eqref{eq:hwc_ddot} (which does not depend directly on $u$) more negative. %The RCBF is again
{\color{black}A fourth possible RCBF{\color{black}, which we denote $H_4^A$,} is then} given by \eqref{eq:predictive_rcbf} with $h_c, u^*_\textrm{orth}$ in place of $h, u^*$, respectively. This choice of $u^*$ is motivated by orbital dynamics, and intuitively, $H^A_4(t_0,x(t_0)) \leq 0$ implies that a safe orbit can be established from $(t_0,x(t_0))$.

Note that we need to verify that $u^*_\textrm{rad}$ and $u^*_\textrm{orth}$ satisfy the requirement of Theorem~\ref{thm:predictive_cbf} that the set $\mathcal{B}(t,x)$ in \eqref{eq:maximizers} contains at most one nonzero element. Assuming a constant $a_\textrm{max} > 0$ as in \eqref{eq:implemented_a_max} exists, it follows that $\ddot{h}_c(t,x,u^*_\textrm{rad}(t,x),w_u) \leq -a_\textrm{max}$, so $\mathcal{B}(t,x)$ in \eqref{eq:maximizers} always has exactly one element{\color{black}, and thus meets the requirements of Corollary~\ref{cor:is_a_rcbf}}. 
% Thus, $u^*_\textrm{rad}$ satisfies the assumption of Theorem~\ref{thm:predictive_cbf} for this system.
\regularversion{\color{black}To meet this criteria, we chose $\rho^A_3 = 2.50(10)^7 \textrm{ m}$. For more complicated control laws such as $u_\textrm{orth}^*$, how to prove that $\mathcal{B}(t,x)$ contains at most one nonzero element is an open research question and is discussed at length in the extended version of this paper \cite{extended}. For this control law, we chose $\rho^A_4 = 476000\textrm{ m}$.}
\extendedversion{However, $u^*_\textrm{rad}$ can still be used even if an $a_\textrm{max}$ does not exist (i.e. if we choose $\rho$ small enough that there exists states $(t,x)\in\mathcal{T}$ such that $\ddot{h}_c(t,x,u^*_\textrm{rad}(t,x),w_u) > 0$). That said, if this is the case, then Theorem~\ref{thm:predictive_cbf} must be applied very carefully. This is because there exists a manifold of initial conditions $(t_a, x_a)$ for which $\chi(t,t_a,x_a), t\geq t_a$ is a closed periodic orbit. Safe trajectories exist on both sides of this manifold, but along this manifold, $\mathcal{B}(t_a,x_a)$ has an infinite number of elements, which violates the assumption of Theorem~\ref{thm:predictive_cbf}. For this reason, we let $\rho^A_3 = 2.50(10)^7 \textrm{ m}$, which places this manifold entirely outside the safe set.}

\extendedversion{
When using $u^*_\textrm{orth}$, there is no constant $a_\textrm{max}$ which bounds $\ddot{h}_c$ (or any higher derivatives), so proving that $\mathcal{B}(t,x)$ in \eqref{eq:maximizers} is nonempty and contains at most one nonzero element is challenging. In practice, many $u^*$, may not allow for straightforward proofs of how many maximizers of $h(t+\beta, \chi(\beta,t,x))$ the system admits. In these cases, how to prove or disprove the applicability of Theorem~\ref{thm:predictive_cbf} is an open research question. For this particular system, our strategy was to examine sample trajectories of $h(t+\beta, \chi(\beta,t,x))$ according to \eqref{eq:ode_with_noise} and 
originating from many different states $(t,x)$. We verified that each trajectory 1) remain bounded, 2) achieved its upper bound, and 3) had at most one nonzero local maximizer. We also sought out corner cases that might violate these three conditions. As no trajectory violating these conditions could be found, we concluded that $u^*_\textrm{orth}$ is consistent with the conditions on $\mathcal{B}(t,x)$ in Theorem~\ref{thm:predictive_cbf}.
We also assume that the initial conditions are such that $\|v_\textrm{orth}\|$ is never zero along the trajectories, so $u^*_\textrm{orth}$ is Lipschitz continuous. Thus, $u^*_\textrm{orth}$ meets the requirements of Theorem~\ref{thm:predictive_cbf} for any $\rho > 0$. For this simulation, we chose $\rho^A_4 = 476000 \textrm{ m}$, which is the radius of Ceres.

In summary, we have four RCBFs $H^A_1, H^A_2, H^A_3, H^A_4$. Each RCBF was constructed using specific properties of the system and the input constraints, so given an initial condition $(t_0,x(t_0))\in\mathcal{S}_H^\textrm{res}$ for any of these RCBFs, we know in advance that there exists at least one safe trajectory beginning from $(t_0,x(t_0))$ along which the control input always satisfies the input constraints. Moreover, the spacecraft will remain on a safe trajectory as long as the control input satisfies the RCBF condition $u \in \boldsymbol\mu_\textrm{rcbf}$ in \eqref{eq:valid_rcbf_control}. Thus, even though the trajectories are not known in advance, we still know the trajectories will stay within the set $\mathcal{S}_H^\textrm{res}$.
}

\subsection{Simulations\extendedversion{: Ceres}} \label{sec:results_ceres}

Next, we validate the RCBFs $H^A_1, H^A_2, H^A_3, H^A_4$ from Section~\ref{sec:safety_setup} in simulation. Suppose a control input of the form
\begin{equation}
    u(t,x) = \argmin_{u \in \boldsymbol\mu_s(t,x,\sigma)} \| u - u_\textrm{nom}(x)\|^2 \,, \label{eq:the_qp}
\end{equation}
where there is a nominal control input $u_\textrm{nom}$ that may be unsafe and $\boldsymbol\mu_s$ is the set of allowable control inputs. In this case, the nominal control input is the linear control law
\begin{equation}
    u_\textrm{nom}\hspace{-1.2pt}(x) \hspace{-2pt}=\hspace{-2pt} - k_p (r - (r \cdot \hat{e}_x)\hat{e}_x ) \hspace{-0.5pt}-\hspace{-0.5pt} k_d \left(v \hspace{-1.2pt}-\hspace{-1.2pt}  \sqrt{\textstyle\frac{2\mu}{\|r\|} \hspace{-2pt}+\hspace{-2pt} 10^4}\hat{e}_x \right) \hspace{-4pt}
\end{equation}
where $k_p = 1.2(10)^{-11}$ and $k_d = 6(10)^{-5}$, and $\hat{e}_x$ is the x-axis unit vector. That is, the nominal control input tries to drive the spacecraft along the x-axis and through the center of Ceres, which is an unsafe state. The set $\boldsymbol \mu_s$ captures the switching described in Section~\ref{sec:switching} and is given by
\begin{equation}
    \boldsymbol{\mu}_s(t,x,\sigma) = \begin{cases} \mathcal{U} & \sigma = 0 \\ \boldsymbol{\mu}_\textrm{rcbf}(t,x) & \sigma = 1 \end{cases} \,.
\end{equation}
The switching tolerances are $\epsilon_1 = 5(10)^4 \textrm{ m}$ and $\epsilon_2 = 1.5(10)^5 \textrm{ m}$, and the class-$\mathcal{K}$ function used to determine $\boldsymbol\mu_\textrm{rcbf}$ was chosen as $\alpha_r$ in \eqref{eq:chosen_alpha}. Note how the quadratic program in \eqref{eq:the_qp} contains only 3 degrees of freedom, and is thus much less computationally expensive than most MPC-based controllers.

\begin{figure}
    \centering
    \includegraphics[width=\columnwidth,trim={0.4in, 0.1in, 0.7in, .3in},clip]{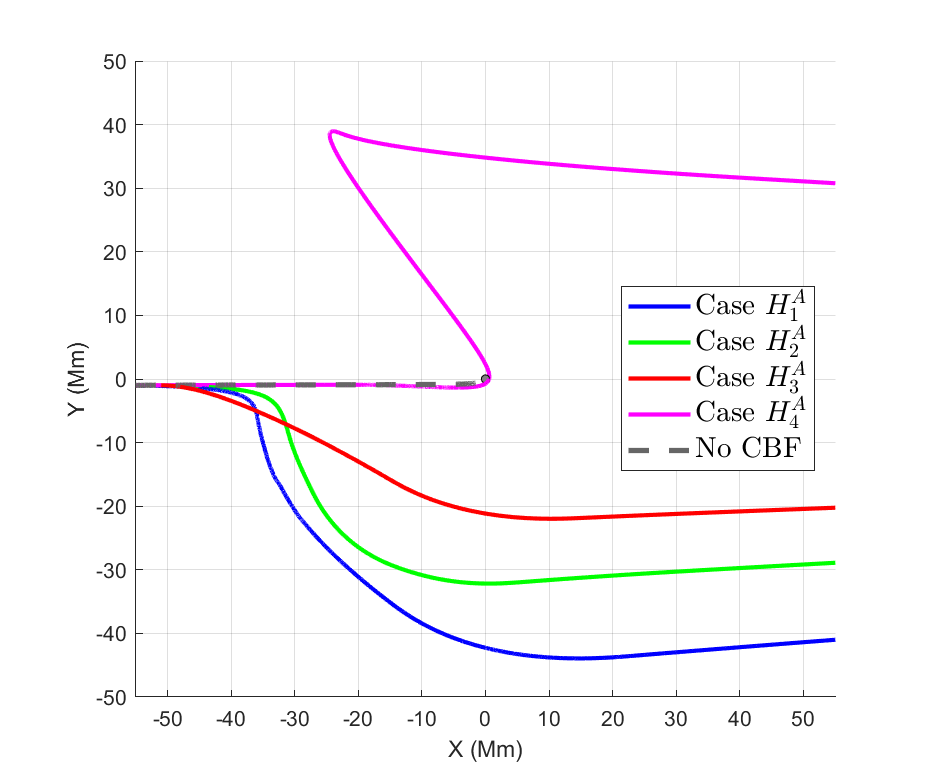}
    \caption{Trajectories under each of the RCBFs {\color{black}moving from left to right}. Ceres is located at the grey dot at the origin and is not to scale. {\color{black}Note how the different RCBFs allow their respective trajectories to approach within differing distances of the asteroid and how the magenta trajectory approaches close enough to be redirected by Ceres' gravity. The grey and magenta trajectories trace very similar paths.}}
    \label{fig:flyby_trajectory}
\end{figure}

We ran four simulations using the above controller with the four different RCBFs $H_1^A, H_2^A, H_3^A, H_4^A$ to specify $\boldsymbol\mu_\textrm{rcbf}$ {\color{black}and one simulation with $\boldsymbol\mu_\textrm{rcbf} = \mathcal{U}$ {\color{black}(i.e. with no RCBF)} for comparison}. All {\color{black}five} simulations had a random {\color{black}zero-mean} matched disturbance of magnitude upper bounded by $w_{u,\textrm{max}} = 5(10)^{-6} \textrm{ m/s}^2$. The simulations using $H_1^A$, $H_2^A$, {\color{black}and no RCBF} also had a random {\color{black}zero-mean} unmatched disturbance of magnitude upper bounded by $w_{x,\textrm{max}} = 2(10)^{-6} \textrm{ m/s}$ (recall that we require $w_{x,\textrm{max}} = 0$ to apply Theorem~\ref{thm:predictive_cbf} for $H_3^A$, $H_4^A$). The initial condition was $x_0 = [-6(10)^7,\,-10^6,\,0,\,20,\,-2,\,0]\transpose$ in all {\color{black}five} simulations, and each was run for 69 simulated days, %(note that multi-month maneuvers are typical when using low-thrust actuators), 
taking on the order of 10 minutes to compute. The resultant trajectories are shown together 
in Fig.~\ref{fig:flyby_trajectory}.

\begin{figure}
    \centering
    \includegraphics[width=0.7\columnwidth,trim={0.4in, 0.1in, 0.7in, .3in},clip]{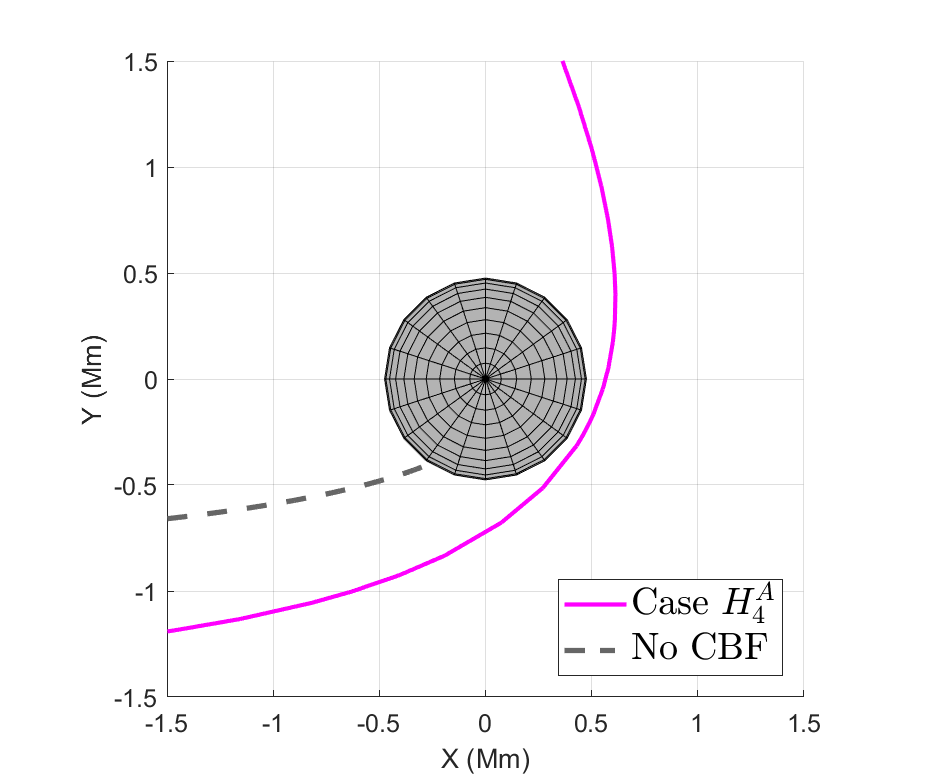}
    \caption{Zoomed-in view of Fig.~\ref{fig:flyby_trajectory} with Ceres to scale {\color{black}(trajectories moving from left to top)}}
    \label{fig:flyby_zoom}
\end{figure}

\begin{figure}
    \centering
    \includegraphics[width=\columnwidth]{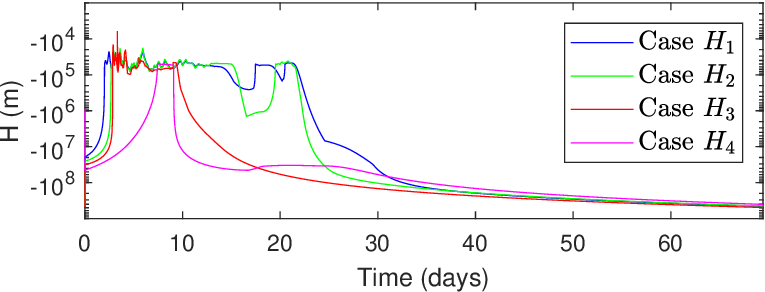}
    \caption{The RCBF values along the trajectories in Fig.~\ref{fig:flyby_trajectory}{\color{black}. Note how each trajectory converges approximately to $H = -\epsilon_1$ due to the choice of $\alpha = \alpha_r$ in \eqref{eq:chosen_alpha}, but the corresponding distances to the asteroid are different in Figs~\ref{fig:flyby_trajectory},\ref{fig:flyby_altitude} due to the differing constructions of the RCBFs.}}
    \label{fig:flyby_cbf}
\end{figure}

\begin{figure}
    \centering
    \includegraphics[width=\columnwidth]{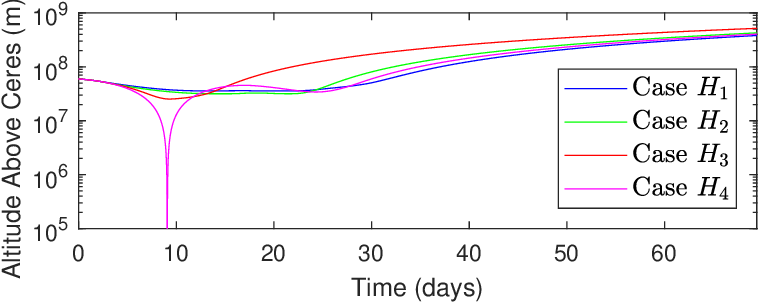}
    \caption{The altitudes above Ceres of the trajectories in Fig.~\ref{fig:flyby_trajectory}}
    \label{fig:flyby_altitude}
\end{figure}

Note how the trajectories under $H^A_1$, $H^A_2$, $H^A_3$ come progressively closer to Ceres because conservatism is reduced with each added layer of complexity. The distance along the x-axis traveled in the same amount of time also increases, as the satellite spends less time reshaping its trajectory around Ceres. 
Physically, the RCBF derivative $\dot{H}$ for $H^A_1, H_2^A, H_3^A$ is minimized when the satellite thrusts away from the asteroid. On the other hand, the RCBF derivative for $H^A_4$ is minimized when the satellite thrusts tangentially to the asteroid. As a result, the trajectory under $H^A_4$ is able to keep $H_4^A$ nonpositive while approaching far closer to Ceres than any of the other RCBFs, as shown in the zoomed-in plot in Fig.~\ref{fig:flyby_zoom}. Here, the satellite comes so close to Ceres that it is redirected by Ceres' gravity (which is much greater than $u_\textrm{max}$), but it is moving fast enough to avoid being pulled outside the safe set{\color{black}, unlike the trajectory with no RCBF, which crashes into the surface of Ceres}. In practice, one might wish to choose a larger $\rho_4^A$ to result in a hyperbola around Ceres with a smaller turn angle than in Fig.~\ref{fig:flyby_zoom}, but the RCBF ensures the trajectory does not impact the asteroid regardless.

\begin{figure}
    \centering
    \includegraphics[width=\columnwidth]{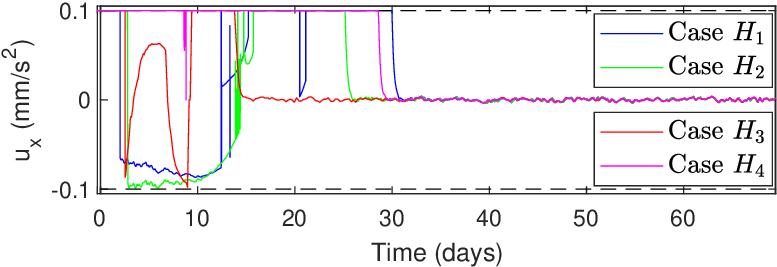}\vspace{6pt}
    \includegraphics[width=\columnwidth]{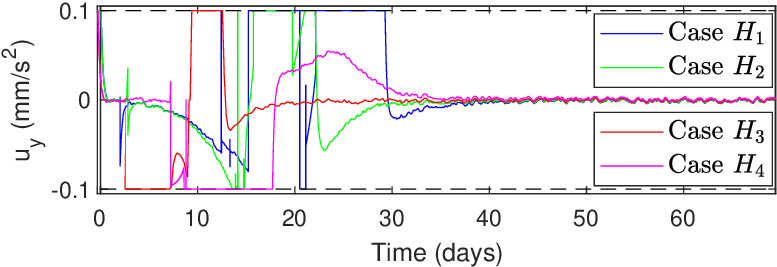}\vspace{6pt}
    \includegraphics[width=\columnwidth]{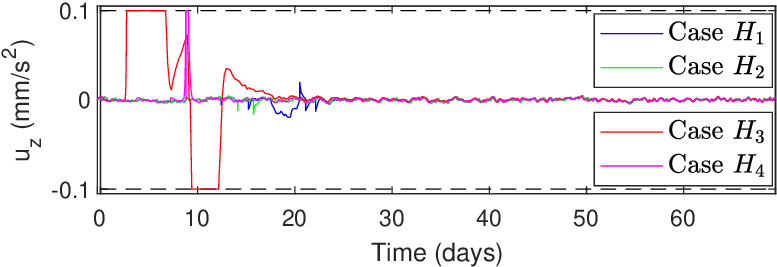}
    \caption{Control inputs along the trajectories in Fig.~\ref{fig:flyby_trajectory}{\color{black}. The controller \eqref{eq:the_qp} remained feasible in the presence of input constraints, as expected from Theorems~\ref{thm:constant_noise_cbf},\ref{thm:variable_cbf},\ref{thm:predictive_cbf}.}}
    \label{fig:flyby_control}
\end{figure}

The values of the RCBFs along these four trajectories are shown in Fig.~\ref{fig:flyby_cbf}.
{\color{black}Note how all four trajectories initially approach the surface $\{x\in\mathcal{S}_H^\textrm{res} \mid H(t,x) = -\epsilon_1\}$ and then remain near this surface until at least $t = 9 \textrm{ days}$.}
Next, the distance to Ceres is plotted in Fig.~\ref{fig:flyby_altitude}. As expected from Fig.~\ref{fig:flyby_trajectory}, the trajectory under $H^A_4$ came the closest to the surface, which may be desirable for an inspection mission, though the satellite was moving very fast during its closest approach. Finally, the control inputs are shown in Fig.~\ref{fig:flyby_control}, and indeed stay within the specified bounds. The z-axis control inputs are negligible, except under $H^A_3$ and $H^A_4$, as $u^*_\textrm{rad}$ and $u^*_\textrm{orth}$ have the potential to magnify the effects of disturbances in the $z$ direction.

If this were a high-thrust scenario (i.e. $u_\textrm{max} > \|f_\mu\|$), any of these four RCBFs would allow the satellite to get equally close to Ceres (in differing time spans), but since the actuators are assumed limited to low-thrust, the choice of RCBF and resultant conservatism made a significant difference in the trajectories. In particular, the RCBF motivated by orbital dynamics, $H^A_4$, allowed for the closest approach to the asteroid. The RCBF $H^A_3$ and its derivatives $\theta, \Theta$ took between 0.25-12 ms to compute, while $H^A_4$ and its derivatives took between 0.36-560 ms to compute on a 3.5 GHz computer, though the code for these calculations could likely be further optimized. In particular, $u^*_\textrm{orth}$ often results in $\chi$ trajectories that travel around Ceres many times, and thus require small time steps to accurately propagate using this state representation. This computation time is also very small relative to time-scale of the problem.

\begin{figure}
    \centering
    \includegraphics[width=\columnwidth]{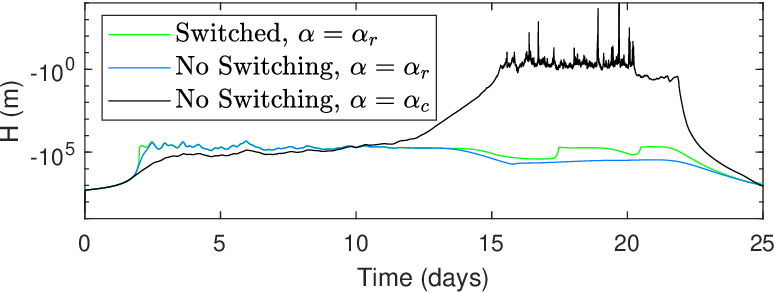}
    \caption{The RCBF values under $H^A_1$ with and without switching and with $\alpha$ both as $\alpha_r$ in \eqref{eq:chosen_alpha} and as a ``regular'' class-$\mathcal{K}$ function $\alpha_c$.}
    \label{fig:flyby_switching}
\end{figure}

Finally, we consider the effect of switching and the choice of function $\alpha$ 
using the RCBF $H^A_1$. First, we ran an additional simulation with $\alpha = \alpha_r$ without switching the RCBF (i.e. $\sigma=1,\forall t\in\mathcal{D}$). The resultant trajectory is qualitatively similar, and a comparison of the RCBF values is shown in Fig.~\ref{fig:flyby_switching}. Here, we see that the trajectory with switching reached {\color{black}the surface $\{x\in\mathcal{S}_H^\textrm{res} \mid H(t,x) = -\epsilon_1\}$} 0.4 days ahead of the trajectory without switching, as expected since the growth rate was not limited by $\alpha_r$ until the satellite came close to the boundary of $\mathcal{S}_{H^A_1}$. 
Second, we ran a simulation with $\alpha$ equal to the ``regular'' class-$\mathcal{K}$ function $\alpha_c$, where $\alpha_c(\lambda) \triangleq k \lambda$, again without switching. We let $k$ equal the average value of $\frac{W}{\epsilon_1}$ when {\color{black}$H_1^A$} was active, which is $k=3.42(10)^{-5}$. Again, the resultant trajectory is qualitatively similar, but we see from Fig.~\ref{fig:flyby_switching} that the two trajectories using $\alpha=\alpha_r$ in \eqref{eq:chosen_alpha} approached $-\epsilon_1$ much faster. However, the trajectory under $\alpha=\alpha_c$ approached much closer to the edge of the safe set than the other two trajectories, because the analysis of the steady state {\color{black}surface $\mathcal{S}_f$} following \eqref{eq:alpha_r_implemented} no longer applies. 

\extendedversion{
\subsection{Simulations: Eros} \label{sec:results_prox} \label{sec:mission_c}

Next, we consider the problem of finding a safe trajectory when the spacecraft is near an irregularly-shaped asteroid, in this case Eros. We consider a mesh model of Eros with $N=3897$ points \cite{shape_model}, shown in 
{\color{black}Fig.~\ref{fig:eros_trajectory}},
where we want the satellite to stay sufficiently far from every point in the mesh. Note that Eros spins with angular velocity $\omega = [3.101(10)^{-4},\,6.232(10)^{-5},\,9.810(10)^{-5}] \textrm{ rad/s}$ \cite{omega_model}, so each point is moving and thus the time varying component $\partial_t H$ is important in this scenario. Denote each point as $r_{c,i},\;i=1,2,\cdots N$. Then we have $N$ time-varying constraint functions
\begin{gather}
    h_{c,i}(t,x) \triangleq \rho - \| r - r_{c,i}(t) \| \,, \\
    r_{c,i}(t) = e^{\omega^\times(t-t_0)}r_{c,i}(t_0) \,,
\end{gather}
where $\omega^\times$ is the cross product matrix for $\omega$.
For this scenario, let $\rho$ be $\rho^B = 500 \textrm{ m}$, which represents the closest allowable distance to any point $r_c$ in the mesh. Note that the maximum distance between points in the model \cite{shape_model} is $850 \textrm{ m}$, so we require $\rho$ to be at least half this distance (otherwise, the agent could travel inside the asteroid between mesh points), and we expect smoother results when $\rho$ and/or $N$ are larger.

We model the gravity of Eros $f_\mu$ using the 16th order spherical harmonics model in \cite{gravity_model}. The full model is assumed known to the controller, but we note that one could also use a simplified model and account for higher order effects as disturbances, as was done in \cite{CDC20}. Since Eros is rotating and asymmetric, $f_\mu$ will also be time-varying. Let $u_\textrm{max} = 0.1 \textrm{ m/s}^2$, which is larger than the peak gravitational acceleration at the surface of Eros, so this is a high-thrust simulation. Suppose there are random matched and unmatched disturbances upper bounded by $w_{u,\textrm{max}}=0.005 \textrm{ m/s}^2$ and $w_{x,\textrm{max}} = 0.001 \textrm{ m/s}$, respectively.

Using these $N$ constraint functions, we construct $N$ RCBFs of the form given in \eqref{eq:constant_noise_cbf}, denoted $H_i^B, i=1,2,\cdots N$. Here, the parameter $a_\textrm{max}$ is the same for every RCBF and is
\begin{equation}
    a_\textrm{max} = u_\textrm{max} - w_{u,\textrm{max}} - \sup_{t\in\mathcal{D}, x\in\mathcal{S}(t)} \| f_\mu(t,r) - u_c(t) \|, \label{eq:eros_a_max}
\end{equation}
where $u_c \neq 0$ since every mesh point is moving. This results in $a_\textrm{max} = 0.0523 \textrm{ m/s}^2$. We could also have found a function $\Phi$ and used the form of RCBF in \eqref{eq:variable_cbf}, but this would be less beneficial here than in Section~\ref{sec:results_ceres} since the agent will always be close to the asteroid surface. We need a separate discrete state for each RCBF to describe whether each RCBF is active, so we introduce the vector $\Sigma$ containing $N$ discrete states $\sigma_i$. Each RCBF induces a set of allowable control inputs $\boldsymbol\mu_{\textrm{rcbf},i}$, so the control inputs must live in the intersection of the allowable sets following from each active RCBF. To this end, define
\begin{equation}
    \boldsymbol\mu_s(t,x,\Sigma) = \bigcap_{\{ i \mid \sigma_i=1\}} \boldsymbol\mu_{\textrm{rcbf},i}(t,x) \,, \label{eq:control_intersection}
\end{equation}
and let $\boldsymbol\mu_s$ equal $\mathcal{U}$ if there is no active RCBF. {\color{black}We assume $\boldsymbol\mu_s$ is always nonempty.} The controller is then
\begin{gather}
    u(t,x) = \argmin_{u \in \boldsymbol\mu_s(t,x,\Sigma)} \| u - u_\textrm{nom} \|^2 \,, \label{eq:eros_qp} \\
    u_\textrm{nom}(x) = -k_p (r - r_t) - k_d v
\end{gather}
where $k_p = 3(10)^{-5}$, $k_d = 0.03$, and $r_t = [20(10)^3,\,0,\,0]\transpose$. Let the switching tolerances be $\epsilon_1 = 100 \textrm{ m}$ and $\epsilon_2 = 300 \textrm{ m}$, and the class-$\mathcal{K}$ functions required to define each $\boldsymbol\mu_{\textrm{rcbf},i}$ are all identical and given by $\alpha_r$ in \eqref{eq:chosen_alpha}.

As observed in \cite{CDC21,Glotfelter2}, we only need to enforce each RCBF condition when it is close to being violated. This is facilitated by the switching approach in Section~\ref{sec:switching} and \eqref{eq:control_intersection}, so the quadratic program in \eqref{eq:eros_qp} never has all 3897 constraints active simultaneously. However, unlike in \cite{Glotfelter2}, the use of hysteresis-switching allows us to achieve this without non-smooth analysis. 
}

\extendedversion{
We then simulated the above controller and all 3897 RCBFs around Eros, starting from initial condition $x_0 = [-20(10)^3,\,-4(10)^3,\,0,\,1,\,1,\,0]\transpose$.
The value of the maximum of the 3897 RCBFs is shown in Fig.~\ref{fig:eros_cbf}. In this simulation, there were never more than 4 RCBFs active simultaneously. The trajectory around the asteroid in an Eros-fixed frame is shown in Fig.~\ref{fig:eros_trajectory} and a video in the inertial frame that also highlights the active RCBFs can be found below\footnote{Animation available at \url{https://youtu.be/ArQ84sdMTqo}}. % Need to update the title frame of the video
The control inputs are shown in Fig.~\ref{fig:eros_control}. As expected, the spacecraft stays safe for all time, despite the natural motion of the asteroid and the nominal linear controller attempting to drive the trajectory through the asteroid. Also, the maximal RCBF value stays very close to $-\epsilon_1$ during the interval $t\in[110, 2760]$ in Fig.~\ref{fig:eros_cbf}.
The choppiness of the control input may be attributed to the sparsity of the mesh relative to the size of the asteroid, and the relatively small values of $\rho$ and $\epsilon_1$.

\begin{figure}
    \centering
    \includegraphics[width=\columnwidth]{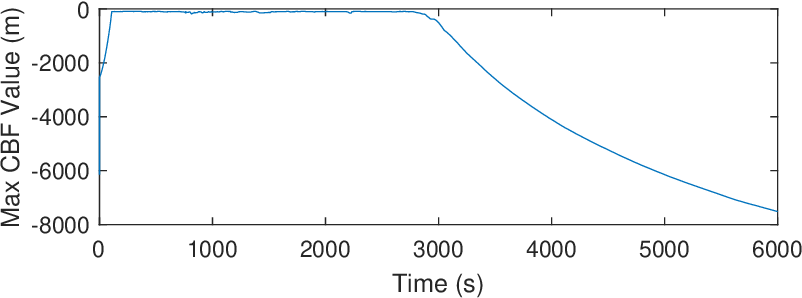}
    \caption{Maximum of the 3897 RCBF values}
    \label{fig:eros_cbf}
\end{figure}

\begin{figure}
    \centering
    \includegraphics[width=\columnwidth,trim={1in, 0.3in, 0.9in, 0.5in}, clip]{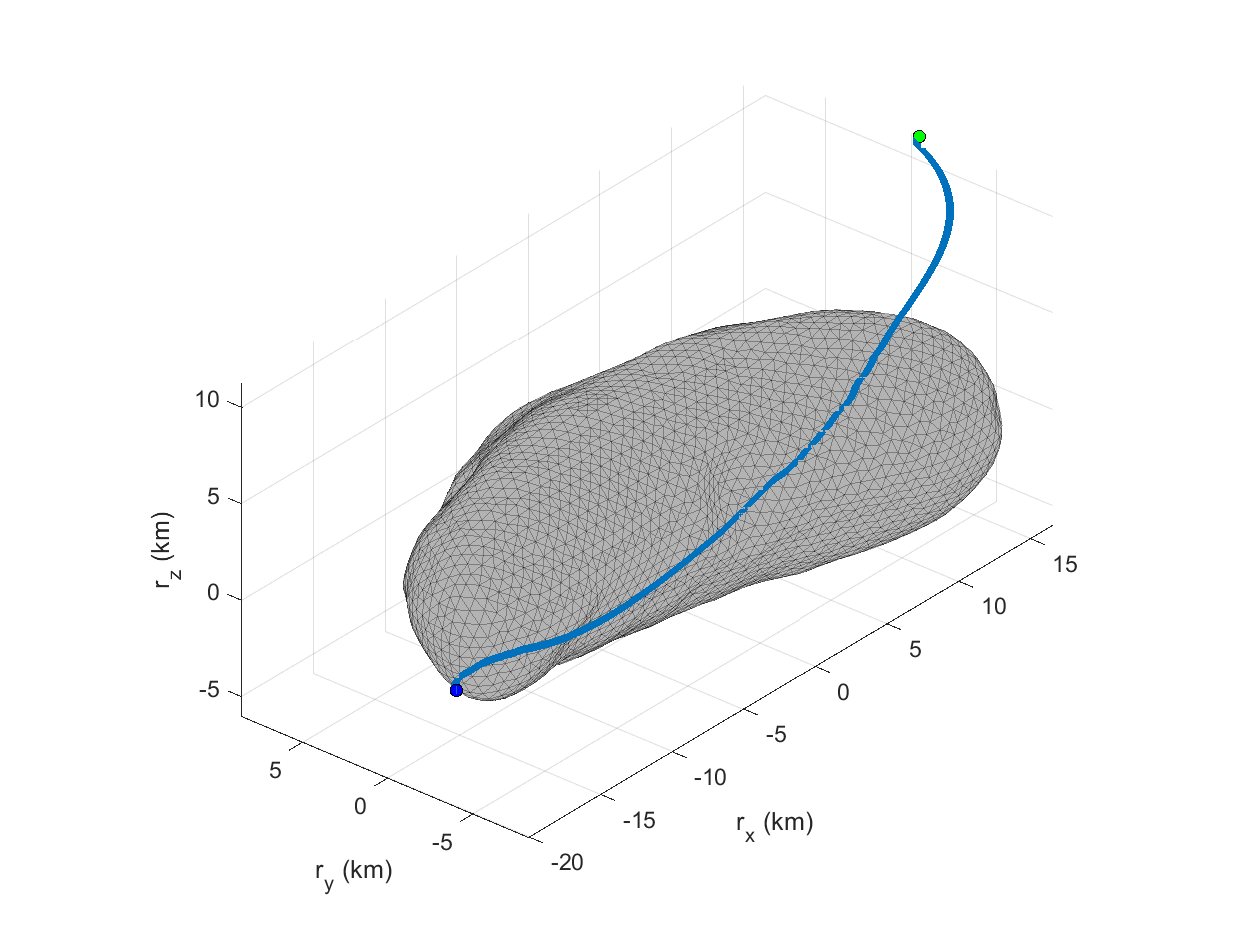}
    \caption{Trajectory of the spacecraft around Eros}
    \label{fig:eros_trajectory}
\end{figure}

\begin{figure}
    \centering
    \includegraphics[width=\columnwidth]{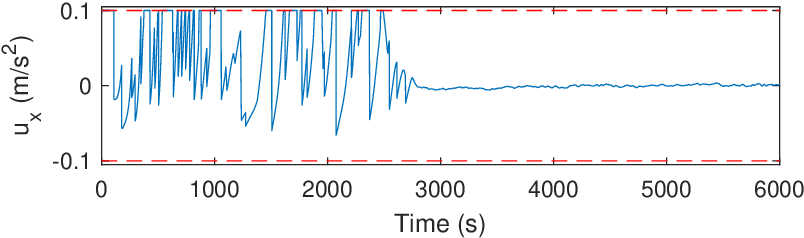}\vspace{6pt}
    \includegraphics[width=\columnwidth]{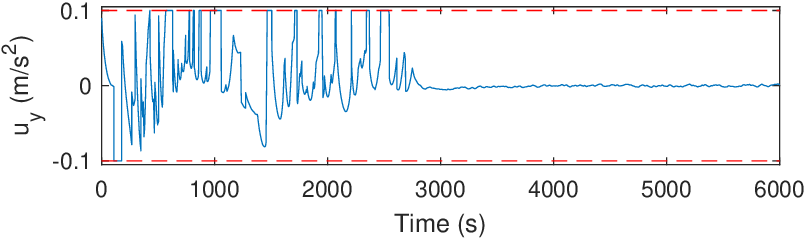}\vspace{6pt}
    \includegraphics[width=\columnwidth]{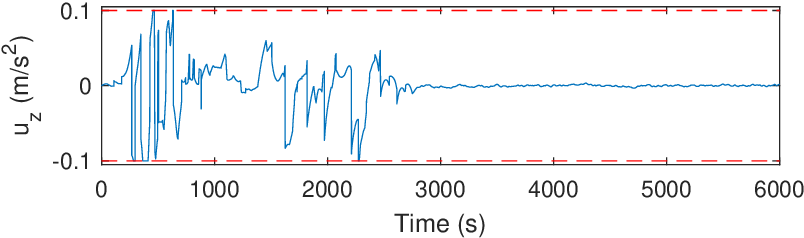}
    \caption{Control inputs of the spacecraft around Eros}
    \label{fig:eros_control}
\end{figure}
}

\section{Conclusions}

We have presented three forms for RCBFs for relative-degree 2 systems that constructively consider input constraints and disturbance bounds to ensure the RCBF condition is always feasible in the presence of input constraints. Thus, systems meeting the theorem requirements are guaranteed to have safe closed-loop trajectories despite the input constraints and disturbances. 
We also introduced a switching approach for enforcing the RCBF condition only near the boundary of the inner safe set, and a class-$\mathcal{K}$-like function that allows us to predict how close the state will approach the boundary of this set as a function of the disturbance. Finally, we applied these methods to create 
\regularversion{\color{black}four}\extendedversion{five}
spacecraft-relevant RCBFs and demonstrated these RCBFs in simulation. 

The simulations show that such RCBFs can be used to plan safe trajectories online, though three of the flyby trajectories were overly conservative. In these simulations, the nominal linear control laws were very simple, so the resultant control inputs were not fuel-efficient. In the future, we are interested in how these RCBFs may be used as constraints in path planning methods for potentially more fuel-efficient trajectories or underactuated systems, and as constraints for scalable multi-agent safe trajectory design. 

\section*{Acknowledgements}

This work was supported by the United States National Science Foundation under the Graduate Research Fellowship Program and under grant no. 1942907.

\bibliographystyle{plain}
\bibliography{sources}

\end{document}